\documentclass[superscriptaddress,footinbib, linenumbers]{ioparxiv}
\usepackage[utf8]{inputenc} 
\usepackage[T1]{fontenc} 
\usepackage{hyperref}       
\usepackage{booktabs}       
\usepackage{nicefrac}       
\usepackage{graphicx}
\usepackage{mathtools}
\usepackage{xcolor}
\usepackage{amsmath}
\usepackage{amssymb, amsthm}
\usepackage{amsfonts}       
\usepackage{physics}
\usepackage{xspace}
\usepackage{bigints}
\usepackage{appendix}
\usepackage{mathtools}
\usepackage{subfig}
\usepackage{paralist}
\usepackage{bm}
\usepackage{xr}
\usepackage{algorithm}
\usepackage{algpseudocode}
\usepackage[algo2e,ruled,vlined,linesnumbered]{algorithm2e}
\usepackage{cleveref}

\makeatletter
\newcommand*{\addFileDependency}[1]{
\typeout{(#1)}
\@addtofilelist{#1}
\IfFileExists{#1}{}{\typeout{No file #1.}}
}\makeatother

\newtheorem{theorem}{Theorem}
\newtheorem{definition}{Definition}
\newtheorem{corollary}{Corollary}
\newtheorem{proposition}{Proposition}
\newtheorem{lemma}{Lemma}

\newtheorem{example}{Example}

\newcommand{\Var}{{\rm Var}}
\newcommand{\Cov}{{\rm Cov}}

\newcommand{\E}[2][]{\ensuremath{\operatorname{\mathbb E}_{#1}\left(#2\right)\xspace}}

\renewcommand{\H}{\ensuremath{\mathcal{H}}}

\newcommand{\Obs}{\ensuremath{\hat{O}}}
\newcommand{\ObsPi}{\ensuremath{\hat{O}_{\mathcal P}}}
\newcommand{\ObsPiU}{\ensuremath{\hat{O}_{\mathcal P, \mathcal U}}}

\newcommand{\Dirichlet}[1][\bm\alpha]{\ensuremath{{\rm Dir}\left(#1\right)}}

\newcommand{\mut}{\ensuremath{\mu_{t}}}
\newcommand{\mutbar}{\ensuremath{\bar{\mu}_{t}}}
\newcommand{\blambda}{\ensuremath{\bm \lambda}}
\newcommand{\bmult}{\ensuremath{\bm m}}
\newcommand{\braketObs}{\ensuremath{\bra\psi \Obs \ket\psi}}
\newcommand{\ExpObs}[1][\Obs]{\ensuremath{\langle #1 \rangle}}
\newcommand{\avgrand}{\ensuremath{\mathcal{D}_t^{(\hat O)}(S)}}
\newcommand{\avgrandperm}{\ensuremath{\mathcal{D}_t^{(\Obs, \Pi)}(S)}}
\newcommand{\avgrandpermU}{\ensuremath{\mathcal{D}_t^{(\Obs, \Pi, U)}(S)}}

\crefname{appendix}{appendix}{appendices}

\graphicspath{ {../figures/} }

\newcommand{\hri}{$ ^1$ Honda Research Institute Europe GmbH, Carl-Legien-Str.\ 30, 63073 Offenbach, Germany}
\newcommand{\aqa}{$ ^2$ $\langle aQa ^L\rangle $ Applied Quantum Algorithms, Universiteit Leiden}
\newcommand{\lorentz}{$ ^3$ Instituut-Lorentz, Universiteit Leiden, Niels Bohrweg 2, 2333 CA Leiden, Netherlands}
\newcommand{\liacs}{$ ^4$ LIACS, Universiteit Leiden, Niels Bohrweg 1, 2333 CA Leiden, Netherlands}

\begin{document}

\title{Hierarchically discriminating Haar-randomness in quantum states from a black-box device}
\author{Xavier Bonet-Monroig $^{1, 2, 3}$, Hao Wang $^{2, 4}$, Adrián Pérez-Salinas $^{2, 3}$}

\affil{\hri}\\
\affil{\aqa}\\
\affil{\lorentz}\\
\affil{\liacs}\\
\email{xavier.bonet@honda-ri.de, h.wang@liacs.leidenuniv.nl, adperez@phys.ethz.ch}

\keywords{Haar-random, $t$-designs, state discrimination, Haar distribution}

\begin{abstract}
The concept of randomness in quantum computing has been central to constructing benchmarking tools, cryptographic protocols, as well as a proof of beyond-classical computation.
Discerning whether quantum states (or unitaries) are randomly distributed is a computational task that requires an enormous amount of quantum computational resources.
This work addresses such a challenge by introducing a hierarchical discrimination algorithm to efficiently test if a set of states $S$ generated from a black-box quantum device with an unknown distribution is (in)compatible with a random distribution.
To this end, we reduce the complexity of the problem by selecting an observable with known spectrum to study the statistical properties of its expectation values with respect to the quantum states from an unknown (black-box) quantum device.
Concurrently, we use our first technical result, a connection between Haar-randomness and the Dirichlet distribution, to analytically compute Haar-random moments of the observable.
Our Haar-random discriminator test is then simply to compare those statistical moments, such that if $S$ fails the test, it is enough to state that the quantum device does not output randomly distributed states.
Else, we can not (yet) confirm that the states follow a Haar-random distribution.
We further provide an extension to this algorithm by permutation- and unitary-equivalent randomization of the observable at increasing computational resources, which allows us to more accurately state whether \(S\) is compatible with Haar-randomness.
We envision the use of the discriminator test as a quantum device benchmark, by discriminating whether the states generated are incompatible with Haar-randomness.
\end{abstract}

\section{Introduction}
Imagine we are given a black-box quantum device that outputs quantum states on demand, and we are asked to efficiently discriminate if their underlying distribution resembles the Haar-random distribution.
Realistically, the black-box might implement randomly chosen sequences of fixed-size quantum circuits, such as a parametrized quantum circuit, or a fixed quantum circuit with random initial states, in all cases possibly without any prior information about their structure.
While this question might be seen just as a thought experiment, the problem of efficiently characterizing the distribution of quantum states is at the core of many problems in quantum information and computation; from quantum advantage and its verification with random circuits or boson sampling~\cite{aaronson2016complexitytheoretic, boixo2018characterizing, aaronson2023certified,arute2019quantum,zhong2020quantum} to cryptography~\cite{ji2018pseudorandom, chen2024power}.
It even plays a central role in the characterization and benchmarking of quantum hardware central \cite{elben2023randomized, eisert2020quantum}, or the performance of variational algorithms~\cite{cerezo2021cost, holmes2022connecting, mcclean2018barren, ragone2023unified}, to mention a few.

As a first step to characterize the output distribution, one might be tempted to take the naive approach: perform full tomography of the quantum states and compare their statistical moments to those of the Haar-random distribution.
Due to the high symmetry of the Haar-random distribution, full tomography quickly faces the curse of dimensionality, and thus requires an exponential cost in quantum resources.

In this work we present an algorithm that initially avoids the exponential cost by hierarchically discriminating the compatibility of the set of states to be Haar randomly distributed.
We base our method on the following observation: a Haar-random distribution must resemble random with respect to all observables, but only one observable showing discrepancies suffices to falsify randomness. 
Our approach aims to hierarchically allocate computational resources for the analysis of expectation values in such a way that the falsification of $t$-designs is possible before facing the unavoidable curse-of-dimensionality.
The first building block of the method is the use of the Dirichlet distribution \cite{johnson1994continuous} to analytically treat expectation values of a known quantum observable w. r. t. Haar-random states as a random variable.
Next, we further extend our method by designing permutation- and unitary-equivalent families of observables to address the true verification of a $t$-design.
By randomly selecting observables in an ordered manner, we discriminate between subspaces of several observables of $t$-designs. 
More importantly, neither this nor any other method suffices to prove that a set of states is Haar-random without facing the curse-of-dimensionality.

This paper is organized as follows. \Cref{sec:background} provides a short background on Haar-randomness. \Cref{sec:Dirichlet} describes the mathematical foundations of this protocol and its link to the Dirichlet distribution. In \Cref{sec:haar} we present the protocol to verify the Haar-randomness of sets of states with respect to a given observable, including extensions for true verification of randomness. We conclude and give an overview of future ideas in \Cref{sec:conclusions}.

\section{Background}\label{sec:background}
Random unitary operations or quantum states are defined via the Haar measure, that is the only measure over the unitary group that remains invariant under the action of an arbitrary operation $U \in \mathcal{SU}(N)$. For unitary operations, the Haar measure $\mu^{(U)}_H$ satisfies
\begin{equation}
    \int_{V \in \mathcal{SU}(N)} d\mu^{(U)}_{H}(V) f(V) = \int_{V \in \mathcal{SU}(N)} d\mu^{(U)}_{H} f(U V), \qquad \forall U \in \mathcal{SU}(N). 
\end{equation}
With respect to states, the Haar-random distribution satisfies
\begin{equation}
    \int_{\substack{\psi \in \mathbb{C}^{N}(N) \\ \braket{\psi}{\psi} = 1}} d\mu_H(\psi) f(\ket\psi) = \int_{\substack{\psi \in \mathbb{C}^{N}(N) \\ \braket{\psi}{\psi} = 1}} d\mu_H(\psi) f(U\ket\psi), \qquad \forall U \in \mathcal{SU}(N). 
\end{equation}

Probability distributions can be studied through their generating functions, or equivalently statistical moments, such as mean and variance.
In the language of Haar-randomness, statistical moments of order $t$ are related to $t$-designs, that is, ensembles of states $S$ satisfying
\begin{equation}
    \rho_t(S) \equiv \E[\ket\psi \sim S]{(\ket\psi\bra\psi)^{\otimes t}} = \int  d\mu_H(\phi) \left(\ket\phi\bra\phi\right)^{\otimes t},
\end{equation}
where the integration limits comprising all possible quantum states have been omitted for readability. 
The generation of $t$-designs is a necessary ingredient to many tasks in quantum computing, such as benchmarking \cite{eisert2020quantum}, and its preparation has been extensively studied.
In particular, $t$-designs can be created with $\mathcal O(\operatorname{poly}(t))$ quantum gates \cite{harrow2023approximate,brandao2016local}, and the minimal number of states required to construct a $t$-design increases as $\Omega(N^t)$ for $t\ll N$ \cite{hughes2021spherical, gross2007evenly}. 

Another interpretation of $t$-designs is obtained by computing expectation values over quantum observables \(\Obs\) such that if a set of states $S$ forms a $t$-design
\begin{equation}\label{eq.eq_moments}
    \mut(\hat O, S) \equiv \E[\psi\sim S]{\bra\psi \hat O\ket\psi^t} = \int  d\mu_H(\phi) \left(\bra\phi\hat O\ket\phi\right)^{t} \equiv \mut(\hat O),\quad \forall \hat{O}
\end{equation}
where the omission of the set of states indicates the Haar-random distribution. 
This condition holds for any quantum observable.
Assume now a \(\Obs\) is given the by rank-1 projector $U\ket x \bra x U^\dagger$, such that $\ket x$ is an element of the computational basis, and $U$ is chosen so that all projectors form an informationally-complete POVM. 
By definition, Haar-randomness induces a strong symmetry across all dimensions of the Hilbert space of interest, and the random variables $\bra x U^\dagger \ket\psi$, for $\ket\psi$ Haar-random, are identically distributed.
Therefore, assessing if \(S\) is a $t$-design implies verifying its compatibility on a set of linearly independent observables of size $\mathcal O(N^2)$.
Yet, a single observable that does not meet \Cref{eq.eq_moments} is sufficient to demonstrate an incompatibility between $S$ and $t$-designs. 
Therefore, verifying Haar-randomness is a much harder task than falsifying it. 
The latter fact is the building block of our Haar-random discriminator algorithm; we aim to efficiently discriminate (e.g. falsify) if a set of quantum states is compatible with \(t\)-designs.

\section{Haar meets Dirichlet}\label{sec:Dirichlet}
The first contribution of this work is to reinterpret the expectation values of Haar-randomly distributed quantum states as random variables with specific distributions.
To this end, we utilize the Dirichlet distribution \cite{johnson1994continuous}, a mathematical tool that allows us to derive an analytical formula for the statistical moments of expectation values $\mut(\Obs)$.
The connection between Haar and Dirichlet has been previously noticed~\cite{boixo2018characterizing}.
For example, the first experiments on quantum advantages base their theoretical separation between classical and quantum computers on this observation, under the commonly used name of Porter-Thomas distribution \cite{arute2019quantum}. 

The intuitive connection between random states and the Dirichlet distribution is straightforward. The Dirichlet distribution ($\Dirichlet$) is a model for multidimensional random variables $\vec x = (x_1, x_2, \ldots, x_m)$ with the constraints that $x_i \in [0, 1]$, and $\sum_i x_i = 1$, and fully determined by a vector of parameters $\vec\alpha$ with the same dimensionality as $\vec x$. That is, the Dirichlet distribution is a probability distribution over the simplex. On the other hand, Haar-random states correspond to states for which $\vert \braket{i}{\psi}\vert^2$ are randomly and uniformly distributed, while the simplex constraint is imposed by normalization. 
A convenient property of the Dirichlet distribution is the existence of a closed form for any statistical moment, for instance $\E[\Dirichlet]{x_i} = \alpha_i / \Vert \bm\alpha\Vert_1$.
We refer the reader to \Cref{app.dirichlet} for details about the Dirichlet distribution. 

\begin{lemma}\label{le.dirichlet}
    Let $\ket\psi$ be a random state drawn from the Haar-random distribution, and let $\bm x$ be a random variable defined as $x_i = \vert \braket{\psi}{i}\vert^2$, for an arbitrary unitary operation $U$, where $\ket i$ is the $i$-th element of the reference basis.
    Then
    \begin{equation}
        \bm x \sim \Dirichlet[\bm 1],
    \end{equation}
    with $\Dirichlet[\bm\alpha]$ being the Dirichlet distribution, and $\bm 1$ being a vector where all entries are $1$.
\end{lemma}
See \Cref{app.dirichlet} for a proof of~\Cref{le.dirichlet}. This result is a direct implication of the identical distributions of random variables $\braket{x}{\psi}$, together with the unitarity constraint of quantum states. 
For completeness, we recall that the marginals of this distribution are all symmetric and more commonly known as the Porter-Thomas distribution~\cite{porter1956fluctuations, boixo2018characterizing} used to build the so-called \textit{quantum supremacy} experiment~\cite{arute2019quantum},
\begin{equation}
    \operatorname{Prob}\left( \left\vert \braket{\psi}{i}\right\vert^2  \ge p\right) = (1 - p)^{N-1} \lesssim \exp(-p N).
\end{equation}

Using this interpretation of Haar-random states, \Cref{le.dirichlet} allows us to analytically compute the expectation value of a Haar-random ensemble of states with respect to an observable $\Obs$. For the sake of simplicity, in the remainder of the paper, we consider $\Obs$ a positive semi-definite Hermitian observable. 
\begin{theorem}\label{th.dirichlet}
Let $\ket\psi$ be a random state drawn from the Haar-random distribution.
Let $\Obs$ be an observable, with $G$ distinct eigenvalues $\blambda = (\lambda_1, \lambda_2, \ldots, \lambda_G), \lambda_i \geq 0$, and algebraic multiplicities $\bmult = (m_1, m_2, \ldots, m_G)$.
Then, the expectation value is a random variable 
\begin{equation}
    \braketObs = \bm \lambda \cdot {\bm x}, 
\end{equation}
where $\bm x$ is a random variable ${\bm x} \sim \Dirichlet[\bmult]$. 
\end{theorem}
To show this, it suffices to observe that the Dirichlet-based description from \Cref{le.dirichlet} holds in any basis, in particular in the one that diagonalizes $\Obs$. Aggregation properties of coordinates of Dirichlet-random variables suffice for this description. A detailed proof can be found in \Cref{app.dirichlet}. 

From the results in~\Cref{th.dirichlet}, it is direct to compute the statistical moments of \(\braketObs\), for Haar-random states.
\begin{lemma}\label{le.moments}
Let $\ket\psi$ be a random state drawn from the Haar-random distribution.
Let $\Obs$ have $G$ eigenvalues $\blambda = (\lambda_1, \lambda_2, \ldots, \lambda_G)$ and multiplicities $\bmult = (m_1, m_2, \ldots, m_G)$.
Then, the $\braketObs$ is a random variable with statistical moments
\begin{equation}\label{eq.exact_moments}
    \mu_t(\Obs) = 
    \sum_{\substack{\bm k \\ \Vert \bm k \Vert_1 = t}} \binom{t}{\bm k} \left(\prod_{i=1}^G \lambda_i^{k_i}\right) \frac{\Gamma(N)}{\Gamma(N + t)} \prod_{i = 1}^G \frac{\Gamma(m_i + k_i)}{\Gamma(m_i)}, 
\end{equation}
with $\Gamma(\cdot)$ being Euler's gamma function. For integers $\Gamma(n+1)=n!$.
\end{lemma}
These computations follow immediately from the multinomial theorem and the properties of Dirichlet distribution.
A detailed proof of this lemma is provided in~\Cref{app.moments}. As an example, consider $\hat O = (I + Z) / 2$, with eigenvalues $\lambda = \{0, 1\}$ and multiplicities $N / 2$. In this case, only the eigenvalues $\lambda \neq 0$ contribute, and we obtain
\begin{align}
    \mu_1(\hat O) & = \frac{1}{2} = \frac{\Gamma(N)}{\Gamma(N + 1)} \frac{\Gamma(N/2 + 1)}{\Gamma(N/2)}\\
    \mu_2(\hat O) & = \frac{1}{4} \frac{N + 2}{N + 1} = \frac{\Gamma(N)}{\Gamma(N + 2)} \frac{\Gamma(N/2 + 2)}{\Gamma(N/2)}
\end{align}

Despite the existence of a closed analytical formula to calculate any \(\mut(\Obs)\), its classical computational cost is prohibitive for large \(t\).
The reason for this bottleneck is the high number of different elements in the multinomial theorem.
Additionally, the rapid growth of the $\Gamma(\cdot)$ function prevents accurate estimations of its values.
Alternatively, it is possible to efficiently estimate \(\mut(\Obs)\) up to precision \(\exp\left(\mathcal O(t^2/N)\right)\) by applying Stirling's approximation to~\Cref{le.moments}.
\begin{corollary}\label{cor.efficient_moments}
    In the assumptions of~\Cref{le.moments}, we can upper- and lower-bound the statistical moments as
    \begin{equation}
    \left(\frac{\Tr \Obs}{N}\right)^t \exp\left( \frac{-t^2}{2N}\right)\leq \mut(\Obs) \leq \left(\frac{\Tr \Obs}{N}\right)^t\exp\left( \frac{t^2}{2 \bar m}\right)
\end{equation} 
with $\bar m^{-1} = \sum_{i = 1} m_i^{-1}$ being an effective degeneracy of the spectrum.
\end{corollary}
A proof of this corollary can be found in~\Cref{app.efficient_moments}.

\section{Hierarchical Haar-random discriminator}\label{sec:haar}
\begin{figure}
    \centering
    \includegraphics[width=0.8\linewidth]{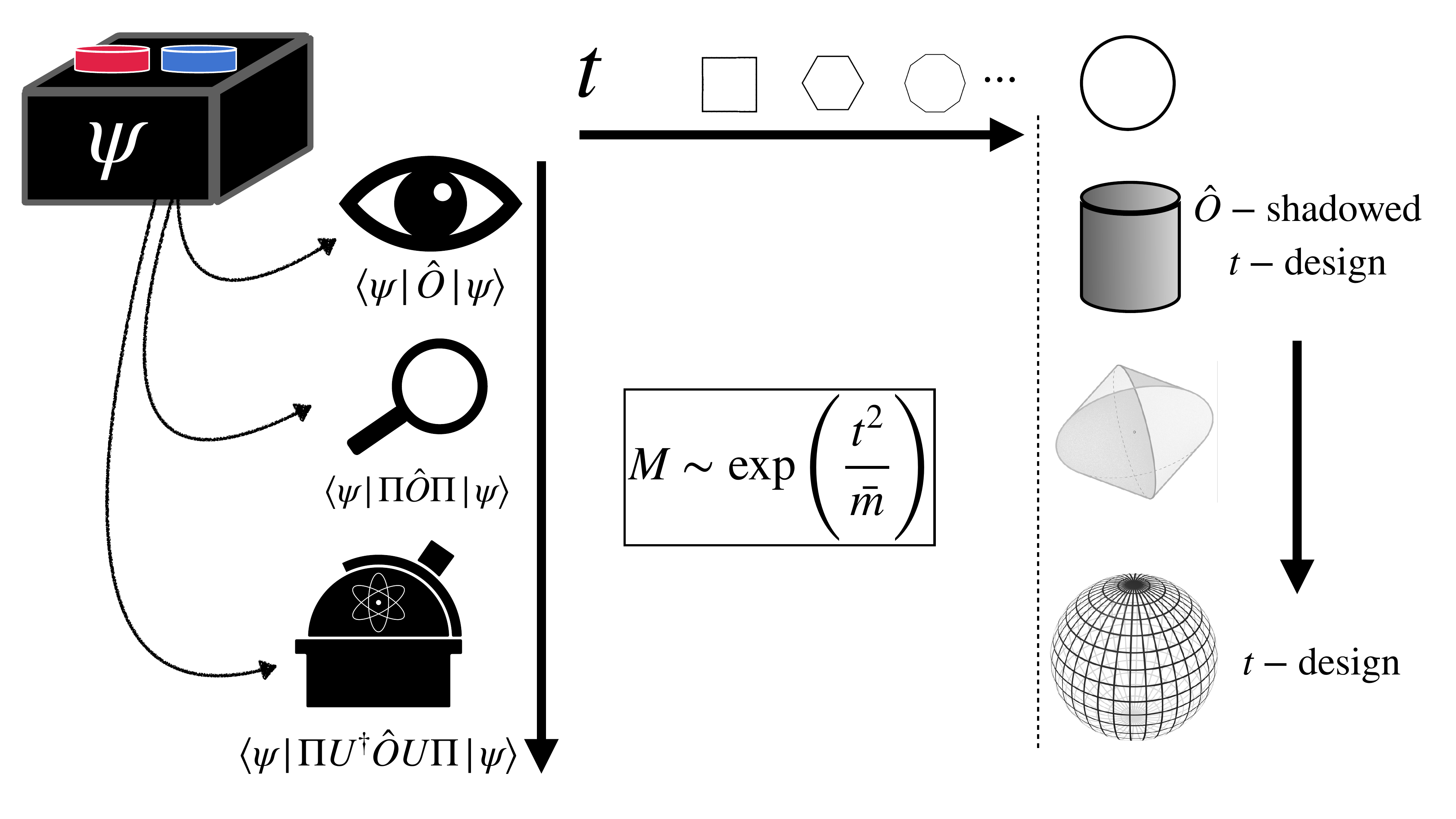}
    \caption{Schematic description of our hierarchical test.
    The goal is to test the randomness of a set of states generated by a black box with two buttons: \textit{copy} and \textit{new state}.
    In this picture, randomness is represented by circles and spheres.
    Our test is built on expectation values of an observable $\hat O$.
    The hierarchy of our test grows in two directions: $t$ (from $t$-designs) and complexity of the observable. 
    An increasing number of $t$ -- polyhedra with many edges -- imposes a denser distribution of states. The sampling cost of the test increases exponentially in $t$. 
    For the observables, $\Obs$ cannot verify $t$-designs due to the high dimension of quantum states. Analogously, the right projection of a cylinder from a sphere are equivalent. 
    To compare states and $t$-designs, we need to construct increasingly more complex families of observables, represented by more sophisticated observation apparatuses.
    The algorithms and subroutines of each of the corresponding schemes are given in \Cref{app.algos}: \(t\)-design check, ~\Cref{alg:tdesigntest}, Permutation check, \Cref{alg:permtest}, Mutually Unbiased Bases check, \Cref{alg:mubtest}, and the combined hierarchical Haar-random discriminator, \Cref{alg:discriminator}.
    }
    \label{fig:scheme_algorithm}
\end{figure}

This section contains our main contribution; an inductive test to discriminate if an unknown set of quantum states $S$ resembles a $t$-design.
The Haar-random discriminator test is described in pseudo-code in~\Cref{app.algos}. 
We strongly encourage the reader to consult the appendix for a comprehensive description of the method.
A visual representation of the Haar-random discriminator algorithm is shown in~\Cref{fig:scheme_algorithm} to help the reader understand it without the need of the algorithmic details.
The starting point is a black box with two buttons to push: produce a copy of the previous state, and produce a new state. 
A given observable $\Obs$ is measured with respect to the states output by the black-box, and the compatibility with a $t$-design according to $\Obs$ is measured, up to a maximum $t$ (top line of \cref{fig:scheme_algorithm}). 
The number of required different random states scales as $M \sim \exp(t^2 \bar m^{-1})$, with $\bar m$ being an effective multiplicity of the eigenvalues of $\Obs$.  
If the test is satisfactory, the observable is extended through permutations, and the procedure to increase $t$ is repeated (middle line of \cref{fig:scheme_algorithm}). 
If the second test is passed, the observable is extended through measurements in other bases (bottom line of \cref{fig:scheme_algorithm}). If this last test does not yield incompatibility, we can only conclude that more tests are needed to verify $t$-designs. If the ensemble is Haar-random (or pseudorandom), our hierarchical test will never falsify the ensemble, for any $t$. 

Our method is said to be hierarchical in the sense that tighter conditions for compatibility with $t$-designs are tested at each step, at more stringent computational resources. 
We give an interpretation in geometrical terms. A Haar-random distribution of states can be interpreted as the uniform distribution over a sphere (over complex numbers). 
A $t$-design requires an even distribution of states to saturate the frame potential \cite{gross2007evenly}. 
The value $t$ captures the degree of the statistical moment to be matched, or equivalently, how fine the even distribution must be \cite{hughes2021spherical, sole1991covering, waldron2003generalized}. 
A direct interpretation is to match a $t$-design with a polygon such that the number of edges grows with $t$, as in \Cref{fig:scheme_algorithm}. 
The observable $\Obs$ is in the end a weighted projector over the Hilbert space. We interpret that as a geometrical projector, in the right column of \Cref{fig:scheme_algorithm}. A $t$-design is a sphere, but there exist bodies whose projections are circles, i. e., indistinguishable from projections of a sphere, such as the cylinder (circular when projected along the $z$ axis) or the bicylinder (circular when projected along $x$ and $y$ axis). Partial projections, such as $\mut(\Obs, S)$ can fool us, hence the need to steadily increase the requirements to verify $t$-designs.   

At the core of the hierarchical discriminator lies the estimation of $\mut(\Obs, S)$ from measurements of the quantum states generated from the black box. 
The device must be able to generate $M$ random and independent quantum states, with $C$ copies of each of them on demand.
We do not require the $C$ copies to be measured simultaneously, depicted in~\Cref{fig:scheme_algorithm} as the red and blue push-buttons.

To estimate $\mut(\Obs, S)$, we compute the unbiased estimator 
\begin{equation}\label{eq:mutbar}
    \mutbar(\Obs, S)= \frac{1}{M}\sum_{\ket\psi \in S} \left(\braketObs\right)^t,
\end{equation}
with numerical error
\begin{equation}\label{eq.mc_error}
    \delta\left(\mutbar(\Obs, S)\right) = \frac{\hat\sigma_t(\Obs, S)}{\sqrt M }.
\end{equation}
The variance $\hat\sigma_t(\Obs, S)$ can be empirically approximated as
\begin{equation}
    \bar\sigma^2_t(\Obs, S) = \frac{1}{M - 1} \sum_{\ket\psi_i \in S} \left((\braketObs)^t - \mutbar(\Obs, S)\right)^2 \approx \bar\mu_{2t}(\hat O, S) -\bar\mu^2_{t}(\hat O, S).
\end{equation}

As we will see, the computing overhead scales exponentially with $t$, making procedures for large values unfeasible.  

We recall that the expectation value $\bra{\psi}\Obs\ket{\psi}$ cannot be exactly determined, hence the need for copies of the same state. The expectation value is then only accessible as an approximation of the form
\begin{equation}
    \bra\psi\Obs\ket\psi^t \mapsto \bra\psi\Obs\ket\psi^t\left( 1 + \frac{\eta}{\sqrt C}\right)^t = \bra\psi\Obs\ket\psi^t \left( 1 + \frac{t\eta}{\sqrt C} + \mathcal O\left( \frac{t^2}{C}\right)\right), 
\end{equation}
with $\eta$ a Gaussian random variable with unit variance, and $C$ the number of copies.
The last equality is given by Taylor expansion.
This implies that the dominant measurement error when estimating $\mut(\Obs, S)$ comes from the number of different states by a factor $C^{-1/2}$.
In the remainder of this manuscript, we neglect the effect of the number of samples by assuming $C \gg t^2$, since typical values of $t$ are small.

\subsection{Increasing $t$-design}\label{sec.tdesign}
The first step for our test is the attempt to falsify the compatibility between \(S\) and a $t$-design measured with respect to a given observable.
For $S$ to be a Haar-random ensemble of states, it must pass the test for arbitrarily large values of $t$. 
Stricter conditions are required as $t$ increases. Notice that $\mu_{t+1}(\Obs, S) = 0 \implies \mut(\Obs, S) = 0$, hence a satisfactory test at degree $t+1$ provides at least the same guarantees as a test for degree $t$. If $S$ is not fully Haar-random, our test will falsify the compatibility between $S$ and a $t$-design for some $t$. 

\begin{example}[Clifford circuits]\label{ex:clifford}
As an example of an ensemble of states for which our hierarchical test works, consider $S_{\rm Stab}$ to be the set of uniformly distributed stabilizer states, i. e., states reached by a random Clifford circuit acting on $\ket 0$. In this case $\avgrand = 0$, up to statistical fluctuations, for $t = \{ 1,2,3\}$ \cite{kueng2015qubit, zhu2017multiqubit}. 
Running the hierarchical test for any \Obs, one might be misled into thinking that $S_{\rm Stab}$ is Haar-randomly distributed if the test is halted in $t\leq 3$.
However, upon inspection of $t=4$, $S_{\rm Stab}$ will fail the average randomness verification test, correctly identifying $S_{\rm Stab}$ not being compatible with a Haar-random distribution of states. 
\end{example}

The compatibility between $S$ and a $t$-design is formally defined via $\Obs$-shadowed $t$-design. 
\begin{definition}[\Obs-shadowed $t$-design]\label{def.oshadow}
    Let $S$ be an ensemble of states, \(\Obs\) a quantum observable, and \(\avgrand\) the \Obs-discriminator of this ensemble such that 
    \begin{equation}\label{eq:Discriminator}
        \avgrand = \mut(\Obs, S) - \mut(\Obs) .
    \end{equation}
    Then, $S$ is a \Obs-shadowed $t$-design, to precision $\epsilon$, if
    \begin{equation}\label{eq:HaarRandDis}
       \left\vert\avgrand\right\vert \leq \epsilon.
    \end{equation}
\end{definition}

To falsify compatibility, it is sufficient to obtain an estimation of \(\avgrand\) such that with high probability
\begin{equation}\label{eq.discriminator_check}
    \vert\avgrand\vert > \delta\left( \avgrand\right),
\end{equation}
and hence we just need to consider a sufficient number of samples to reduce the error below this threshold. 
\begin{theorem}[Samples to falsify $\Obs$-shadowed $t$-designs]\label{th.Haar_check}
Let $S$ be a set of \(N\)-dimensional states.
A number of different states $M$ suffices to falsify the compatibility of $S$ with a \Obs-shadowed $t$-design, with
\begin{equation}\label{eq.montecarlosamples}
    M \geq \frac{\exp\left(\frac{t^2}{2 \bar m}\right)}{\epsilon^2}\left( \frac{\Tr(\Obs)}{N}\right)^{2t}, 
\end{equation}
with $\bar m^{-1} = \sum_{i = 1}^G m_i^{-1}$, and $m_i$ the multiplicities of the eigenvalues of $\hat O$.
\end{theorem}
A proof of the theorem can be found in~\Cref{app.Haar_check}. The scaling of $M$ dominates the cost of the randomness verification, being $\mathcal O\left(\exp\left(\frac{t^2}{\bar m}\right)\right)$ in relative error.

In this analysis, we assume knowledge of the spectrum of \(\Obs\) to compute the analytical values $\mut(\Obs)$.
One might be tempted to optimize the eigenvalue distribution to tune the obtained values of $\mut(\Obs)$, and the number of Monte Carlo samples required to verify $\Obs$-shadowed Haar-randomness to a certain precision.
Optimal spectral distributions must exist to minimize this bound. Additionally, sets of (non-commuting) observables $\{\Obs_1, \Obs_2, \cdots\}$ can be investigated. From \Cref{eq.eq_moments} it is clear that product of expectation values, e. g. $\bra\psi\Obs_1\ket\psi \bra\psi\Obs_2\ket\psi$ must also match their distributions on Haar-random states.
The above procedure can be extended at the expense of classical post-processing via numerical estimation of high-order moments.
These two questions are left for future research.

Thus far, we have been able to discern whether \(S\) is a $\Obs$-shadowed $t$-design, but this is not sufficient to unequivocally disprove that \(S\) is Haar-random.
In our protocol, randomness is lost at two points:
\begin{inparaenum}[(i)]
\item The aggregation properties used in~\Cref{th.dirichlet} imply that measurements via $\Obs$ are only capable of distinguishing between eigenspaces, but any properties of the states within the subspace are ignored.
\item \Cref{le.dirichlet} implies that any measurement is sensitive to the amplitudes of the coefficients in the eigenbasis of $\Obs$, yet the relative phases of these quantum states are completely neglected.
\end{inparaenum}
In the rest of the manuscript we provide means to overcome these limitations, and its computational cost.

\subsection{Permutations}\label{sec.perms}
The first step towards improving the Haar-random discrimination capabilities is to ensure that states within the same subspace are also tested.
This can be done by randomizing the spectrum of the \(\Obs\) by applying permutations.
Furthermore, this can be achieved without the need of additional quantum resources, re-using the already measured expectation values in classical post-processing.
This step corresponds to going from the naked eye to a magnifying glass in~\Cref{fig:scheme_algorithm}, a slight increase in cost for a significant improvement in visual capacity.

\begin{example}[Relevance of permutations]
Consider the family of random states
\begin{equation}
    \ket\psi = \sum_{k = 0}^n \sqrt{p_k} \ket 0^{\otimes k} \ket 1^{\otimes n - k},
\end{equation}
with $\bm p = \{p_1, \ldots, p_n\}$ sampled from a Dirichlet distribution defined by the parameters $\bm\alpha_k = \binom{n}{k}$, for all $k \in \{0, 1, \ldots, n \}$.
To discriminate the Haar-randomness of the ensemble, let us consider the observable
\begin{equation}\label{eq:observable_Z}
    \Obs = \sum_{q= 1}^n \frac{1 - Z_q}{2},
\end{equation}
with $Z_q$ being the $Z$-Pauli matrix acting on the $q$-th qubit.
This observable is diagonal in the computational basis, $\Obs \ket k = \lambda_k \ket k$, with eigenvalues $\lambda_k$ being the number of $1$s in the binary representation of the state.
The spectral distribution of $\Obs$ is $\lambda = \{0, 1, \ldots n\}$, with multiplicity $\binom{n}{k}$.
In this example, the set of states is specifically designed to accumulate all the relative sizes of the eigenspaces (defined by $\Obs$) in a one-dimensional linear subspace.
One can see that such a set of quantum states is not random under the Haar-random measure, yet it appears to be when measured with respect to \(\Obs\).
By simply applying a unitary transformation to the states \(\Obs\)-shadowed is no longer Haar-random, showcasing the limitations of the protocol in \Cref{sec:haar}.
\end{example}

To overcome the hurdle, we can average over permutations on the spectrum of \(\Obs\) via Monte Carlo, a process that requires only classical post-processing.
We extend~\Cref{def.oshadow} to the family of observables defined as $\Obs_\Pi = \Pi \Obs \Pi$, where $\Pi$ is any permutations of $N$ elements in the eigenbasis of $\Obs$.
We define the set $S$ being a $(\ObsPi)$-shadowed $t$-design if 
\begin{align}
    \avgrandperm = &
    \E[\Pi]{\mut(\Obs_\Pi, S)} - \mut(\Obs) \\ 
    \left\vert\avgrandperm\right\vert < & \epsilon,
\end{align}
and we extend~\Cref{th.Haar_check} to include permutations.

\begin{corollary}\label{cor.permutation}
Let $\mathcal P$ be a set of permutations.
Fix a permutation $\Pi$.
For this $\Pi$ we estimate $\avgrandperm$ via Monte Carlo with $M$ samples.
We repeat the process with $M_\Pi$ different permutations.
We can verify that $S$ is a $(\ObsPi)$-shadowed $t$-design if 
\begin{equation}
    \left\vert\avgrandperm\right\vert \leq \left(\frac{\Tr\Obs}{N}\right)^t \exp\left( \frac{t^2}{4 \bar m}\right) \left(\frac{1}{\sqrt{M_\Pi}} + \frac{1}{\sqrt{M}}\right)
\end{equation}
\end{corollary}
The proof is a standard error propagation of Monte Carlo errors. 

Now, we connect the arguments of the Dirichlet distribution with the permutations $\Pi$.
The action of $\Pi$, as seen from the Dirichlet distribution from which $\bm x$ is sampled, is to permute the coefficients $\bm\alpha$.
Consider the case where $S$ is a $(\Pi\Obs\Pi)$-shadowed $t$-design for some $\Pi$.
Then, a subset of $\bm\alpha$ need not be affected by such permutation, and subsequently, the underlying Dirichlet distribution must be symmetric.
We can argue that if $\avgrand = \mathcal R_t^{(\Pi \Obs \Pi)}(S) \approx 0$ for sufficiently many permutations $\Pi$, then exchanging any set of coefficients leads to the same random variable, and in turn, implies a fully symmetric Dirichlet distribution as in~\Cref{le.dirichlet}.

One can compute \(\mut(\Obs, S)\) by solving a linear system of equations with the partial statistical moments {\scriptsize \( \E{\prod_{i} x_i^{k_i}}\)} as the unknown variables.
Clearly, this vector has exponential size, and thus one must run over exponentially many permutations to solve the linear system.
Alternatively, we can leverage our sampling strategy to statistically estimate how close a set of states is from the symmetric Dirichlet distribution, which is invariant under permutation.
\begin{proposition}[Non-symmetry of Dirichlet distribution]\label{prop.nonsymmetric}
Let $S$ be an ensemble whose sampling provides underlying probability distributions of the form $\Dirichlet$, with unknown $\bm\alpha$. Then, the estimation of $\mut(\ObsPi, S)$ can be used to estimate the average deviations between the coordinates of $\bm\alpha$ by
\begin{equation}
    \E[i, j]{(\alpha_i - \alpha_j)^2} \in \mathcal O\left( \left(\frac{\Tr(\Obs^2)}{N^2}\right)^{-t} \Var_{\Pi}\left( \mu_t(\ObsPi, S) - \mut(\Obs)\right)\right). 
\end{equation}
\end{proposition}
This proposition is not a formal proof, but rather an educated guess to estimate the average values. The full derivation is available in~\Cref{app.nonsymmetric}.

\subsection{Mutually Unbiased Bases}\label{sec.mub}
Even after one has sufficiently permuted over the eigenbasis of \(\Obs\) without signaling any incompatibility with Haar randomness, it is not sufficient to guarantee that the states form a \(t\)-design.
The reason for such insufficiency lies in the fact that the information retrieval of measuring \(\braketObs\) is necessarily incomplete.
By simple counting, we observe that permutations limit the number of accessible degrees of freedom to \(2^n\).
Yet, full tomography of a \(n\)-qubit quantum state requires at least \(2^{2n}\) independent measurements~\cite{banaszek2013focus}.

To illustrate the relevance of the measurement basis, we recall an example based on separations between classical and quantum computers.
\begin{example}[Relevance of measurement basis]\label{ex:relevance_basis}
Instantaneous Quantum Polynomial (IQP) circuits are defined as 
\begin{equation}
    U_{\rm IQP} = H^{\otimes n} D H^{\otimes n}, 
\end{equation}
for $D$ a non-Clifford diagonal gate, that is with $T$ or $CCZ$ gates.
The task of sampling from these circuits cannot be simulated by any classical algorithm (under commonly accepted conjectures in computational complexity) \cite{bremner2016averagecase}, and it has been observed that its sampling has a similar underlying distribution to that of random circuits \cite{boixo2018characterizing}.
Consider again the observable \Cref{eq:observable_Z}, then the family of IQP states seems Haar-random.
Permutations of the computational basis do not significantly modify this state, since they can be absorbed into $D$.
However, IQP circuits cannot be Haar-random due to a lack of degrees of freedom (e.g. sampling in the $X$ basis returns uniformly distributed bitstrings).
In this case, sampling in the $X$ basis suffices to falsify the Haar-randomness of IQP circuits.  
\end{example}

Our next step is to show that the Haar-random hierarchical test can be extended to an informationally complete set of measurements.
This additional step allows us to properly test compatibility with $t$-designs, going beyond the its shadowed version.
To achieve information completeness, we resort to complete sets of mutually unbiased bases (MUB) \cite{scott2006tight}, only requiring Clifford circuits \cite{kern2010complete}. 
Then, we only need to transform $\Obs$ through a quantum circuit and repeat the previous steps in the algorithm (see~\Cref{app.algos} for the pseudo-code description).

\begin{lemma}[]\label{le.ic_povm}
Let $S$ be a set of states of dimension $N$.
Let $\Obs$ be an observable, let $\mathcal P$ be the set of all permutations over the eigenbasis of $\Obs$, and let $\mathcal U$ be a complete set of MUB.
Then, the set of operators $\Obs_{\Pi,U} = \{\Pi U^\dagger \Obs U \Pi\}$ for $\Pi\in\mathcal P, U \in \mathcal U$ is tomographically complete. 
\end{lemma}
The operators of the form $\Pi_1 \Obs \Pi_1 - \Pi_2\Obs\Pi_2$, can be used to construct observables as $\Vert \Obs \Vert \left(\ket i \bra i - \ket j \bra j\right)$, for any pair $(i, j)$. 
These operators, plus the identity, suffice to obtain all relevant information in the eigenbasis of $\Obs$, which combined with the set of MUB, are sufficient to construct a tomographically complete measurement~\cite{scott2006tight}.
See~\Cref{app.ic_povm} for a detailed proof.

This observation motivates the use of MUB for extending $\Obs$-shadowed Haar-randomness.
We define a set of states $S$ to be $\epsilon$-close to a $(\ObsPiU)$-shadowed $t$-design if 
    \begin{align}
        \avgrandpermU & =
        \E[\Pi, U]{\mut(\Obs_{\Pi,U}, S)} - \mut(\Obs)  \\ 
        \left\vert\avgrandpermU\right\vert &<  \epsilon,
    \end{align}
where $\mathcal U$ is complete set of MUB and $\mathcal P$ is the set of all permutations in the eigenbasis of $\Obs$. Notice that $U$ can absorb the permutations. We choose to keep permutations separately to lower the demands on quantum resources.  

As in previous sections, we can estimate the statistical moments $\mutbar(\Obs_{\Pi,U})$ via Monte Carlo to approximately obtain \(\avgrandpermU\).
This allows us to statistically verify the Haar-randomness of the set of states, irrespective of the observable $\Obs$. 
\begin{corollary}\label{cor.MUB}
Let $S$ be a set of states of dimension $N$.
Let $\Obs$ be a $N\times N$ Hermitian matrix, and let $\{ \Pi\}$ be the set of all permutations, and let $\mathcal U$ be a complete set of MUB.
Fix a basis $U \in \mathcal U$.
For this $U$ we estimate ${\mathcal{R}^{(U^\dagger \Pi \Obs \Pi U)}_t(S)}$ via Monte Carlo, with $M_\Pi$ different permutations, each with $M$ samples, and repeat the process with $M_U$ different bases.
We can verify that $S$ is compatible with a $(\ObsPiU)$-shadowed $t$-design if 
\begin{equation}
    \left\vert\avgrandpermU\right\vert \leq \left(\frac{\Tr\Obs}{N}\right)^t \exp\left( \frac{t^2}{4 \bar m}\right)\left(\frac{1}{\sqrt{M}} + \frac{1}{\sqrt{M_\Pi}} + \frac{1}{\sqrt{M_U}}\right)
\end{equation}
\end{corollary}
The tomographical completeness of the measurements stated in~\Cref{le.ic_povm}, implies that low values of $\avgrandpermU$ are only compatible with $S$ being approximately a $t$-design. 
This step inevitably requires additional quantum resources that potentially grow exponentially with the order of the \(t\).
In our pictorial representation this step corresponds to going from a magnifying glass to a radio telescope, a simile that perfectly captures the essence of our method; gaining immense understanding of our surroundings at a very high cost.

\subsection{Mixed states and frame potential}
Finally, we can go beyond pure states and extend the algorithm to a set \(S\) of mixed states.
In the Schmidt decomposition,  $\rho$ is given by
\begin{equation}
\rho = \sum_{i = 1}^N p_i U \ket{i} \bra{i} U^\dagger. 
\end{equation}
The set $S$ is now defined by $p_i$ and $U$, where both elements are drawn from two different probability distributions.
Consider the case in which $U$ is drawn from the unitary Haar distribution, then we can reabsorb any matrix diagonalizing \(\Obs\) into $U$.
Analogously to~\Cref{th.dirichlet}, the expectation value is a random variable specified by
\begin{equation}
    \Tr\left( \rho \ \Obs \right) = \sum_{i, j} p_i \lambda_j \left\vert \bra{j} U \ket i \right\vert^2.
\end{equation}
Random matrices are accessible by applying the QR decomposition to an $N \times N$ matrix filled with random Gaussian variables, or by sampling random $N$-dimensional vectors from spaces of decreasing dimensions~\cite{meckes2019random}.
Unfortunately, $\left\vert \bra{j} U \ket i \right\vert^2$ does not admit a simple representation as in the case of pure states~\cite{zyczkowski1994random}.
A detailed description of $\Tr\left( \rho \Obs \right)$ variable is out of the scope of this work and left for future research.

Another connection of the Haar-random discriminator algorithm is that of frame potentials~\cite{hunter-jones2019unitary, gross2007evenly}.
A Monte Carlo procedure sampling relative overlaps between states from an unknown set allows us to numerically estimate the frame potential, which can be compared to existing analytical bounds~\cite{welch1974lower}.
This procedure will yield results similar to those of this manuscript with different requirements.
In the case of frame potentials one would require simultaneous copies of different states in \(S\), as well as SWAP operations for a fidelity test.
In contrast, our hierarchical discriminator algorithm only requires measurements of individual expectation values and classical post-processing.
A very interesting future research question is to compare which of these two approaches yields more robust results with the same quantum resources.

\section{Conclusions}\label{sec:conclusions}
We have introduced a hierarchical discriminator to falsify the compatibility between a set of states $S$ and a $\Obs$-shadowed $t$-design, i.e. a $t$-designed projected on a fixed observable $\Obs$. 
The test is hierarchical in two directions:
larger $t$ values impose an exponentially-in-$t$ number of evenly distributed states in $S$, as seen by a fixed \(\Obs\).
Second, the observable $\Obs$ can be modified via random permutations and change of bases to extend the test from $\Obs$-shadowed to actual $t$-designs in a statistical sense.
The input of this method requires only the expectation values of the observable \(\Obs\).
Our first main result is to show that the expectation values of an observable measured with Haar-random states output values that can be related to a Dirichlet distribution.
The second main result shows that if the spectrum of the observable is known, then the statistical moments of the expectation values of Haar-random states can be calculated exactly, albeit computationally costly.
To overcome this cost, we provide an efficient method to lower- and upper-bound the Haar-random moments.
We can extend these results by artificially increasing the measurements on the ensemble.
Permutations in the eigenbasis of $\Obs$ allow us to verify the symmetry of the underlying Dirichlet distributions.
Mutually unbiased bases allow us to perform tomographically complete measurements.
At the end of the procedure, if the measurements are on average compatible with a \(t\)-design, then the ensemble of states must also be (approximately) a \(t\)-design.

We foresee applications of the discriminator method presented here in multiple directions.
First, it provides a rapid and efficient Haar-randomness check, potentially preventing the squandering of quantum computational resources.
In addition, a test of \(\Obs\)-shadowed $t$-designs with respect to a single observable might be useful in the context of trainability of parameterized quantum circuits where an observable \(\Obs\) is used to define a cost function.
Our Haar-random discriminator method will indicate the presence of vanishing gradients or barren plateaus~\cite{mcclean2018barren}, or the lack thereof as hinted in~\Cref{sec:haar}.
Our formulation clearly exposes that $t$-designs are sufficient yet not necessary for them to arise.
Moreover, this method can be used as an offline characterization tool for variational algorithms to explore heuristic approaches in areas such as optimization or quantum machine learning, in alignment with recent research~\cite{perez-salinas2024analyzing}.
Finally, the hierarchical discriminator protocol can also serve as a statistical alternative to tomographical verification of spherical $t$-designs if $\avgrandpermU \approx 0$, under the condition that $t \in \mathcal O(1)$ to be computationally affordable.
One can envision the creation of a benchmark metric to characterize quantum hardware quality by how accurate it generates \(t\)-design quantum states~\cite{harrow2023approximate}.
 
\section*{Acknowledgements}
The authors would like to thank Carlo Beenakker, Jordi Tura-Brugu{\'e}s and Vedran Dunjko for their support on this project, and Andrei Udriste, Stefano Polla, Patrick Emonts, Tim Coopmans and Berta Casas for useful comments.
The authors would like to extend their gratitude to all members of aQa Leiden for fruitful discussions.
This work was supported by the Dutch National Growth Fund (NGF), as part of the Quantum Delta NL programme.

\bibliographystyle{iopart-num}
\bibliography{references}

@misc{aaronson2016complexitytheoretic,
  title = {Complexity-{{Theoretic Foundations}} of {{Quantum Supremacy Experiments}}},
  author = {Aaronson, Scott and Chen, Lijie},
  year = 2016,
  month = dec,
  number = {arXiv:1612.05903},
  eprint = {1612.05903},
  primaryclass = {quant-ph},
  publisher = {arXiv},
  doi = {10.48550/arXiv.1612.05903},
  urldate = {2023-04-03},
  archiveprefix = {arXiv}
}

@misc{aaronson2023certified,
  title = {Certified {{Randomness}} from {{Quantum Supremacy}}},
  author = {Aaronson, Scott and Hung, Shih-Han},
  year = 2023,
  publisher = {arXiv},
  doi = {10.48550/ARXIV.2303.01625},
  urldate = {2026-02-10},
  copyright = {arXiv.org perpetual, non-exclusive license}
}

@misc{appleby2014galois,
  title = {Galois {{Unitaries}}, {{Mutually Unbiased Bases}}, and {{MUB-balanced}} States},
  author = {Appleby, D. M. and Bengtsson, Ingemar and Dang, Hoan Bui},
  year = 2014,
  month = oct,
  number = {arXiv:1409.7987},
  eprint = {1409.7987},
  primaryclass = {quant-ph},
  publisher = {arXiv},
  doi = {10.48550/arXiv.1409.7987},
  urldate = {2023-05-31},
  archiveprefix = {arXiv}
}

@article{arute2019quantum,
  title = {Quantum Supremacy Using a Programmable Superconducting Processor},
  author = {Arute, Frank and Arya, Kunal and Babbush, Ryan and Bacon, Dave and Bardin, Joseph C. and Barends, Rami and Biswas, Rupak and Boixo, Sergio and Brandao, Fernando G. S. L. and Buell, David A. and Burkett, Brian and Chen, Yu and Chen, Zijun and Chiaro, Ben and Collins, Roberto and Courtney, William and Dunsworth, Andrew and Farhi, Edward and Foxen, Brooks and Fowler, Austin and Gidney, Craig and Giustina, Marissa and Graff, Rob and Guerin, Keith and Habegger, Steve and Harrigan, Matthew P. and Hartmann, Michael J. and Ho, Alan and Hoffmann, Markus and Huang, Trent and Humble, Travis S. and Isakov, Sergei V. and Jeffrey, Evan and Jiang, Zhang and Kafri, Dvir and Kechedzhi, Kostyantyn and Kelly, Julian and Klimov, Paul V. and Knysh, Sergey and Korotkov, Alexander and Kostritsa, Fedor and Landhuis, David and Lindmark, Mike and Lucero, Erik and Lyakh, Dmitry and Mandr{\`a}, Salvatore and McClean, Jarrod R. and McEwen, Matthew and Megrant, Anthony and Mi, Xiao and Michielsen, Kristel and Mohseni, Masoud and Mutus, Josh and Naaman, Ofer and Neeley, Matthew and Neill, Charles and Niu, Murphy Yuezhen and Ostby, Eric and Petukhov, Andre and Platt, John C. and Quintana, Chris and Rieffel, Eleanor G. and Roushan, Pedram and Rubin, Nicholas C. and Sank, Daniel and Satzinger, Kevin J. and Smelyanskiy, Vadim and Sung, Kevin J. and Trevithick, Matthew D. and Vainsencher, Amit and Villalonga, Benjamin and White, Theodore and Yao, Z. Jamie and Yeh, Ping and Zalcman, Adam and Neven, Hartmut and Martinis, John M.},
  year = 2019,
  month = oct,
  journal = {Nature},
  volume = {574},
  number = {7779},
  pages = {505--510},
  publisher = {Nature Publishing Group},
  issn = {1476-4687},
  doi = {10.1038/s41586-019-1666-5},
  urldate = {2021-05-18},
  copyright = {2019 The Author(s), under exclusive licence to Springer Nature Limited},
  langid = {english}
}

@article{bailey1992distributional,
  title = {Distributional {{Identities}} of {{Beta}} and {{Chi-Squared Variates}}: {{A Geometrical Interpretation}}},
  shorttitle = {Distributional {{Identities}} of {{Beta}} and {{Chi-Squared Variates}}},
  author = {Bailey, Ralph W.},
  year = 1992,
  journal = {The American Statistician},
  volume = {46},
  number = {2},
  eprint = {2684178},
  eprinttype = {jstor},
  pages = {117--120},
  publisher = {[American Statistical Association, Taylor \& Francis, Ltd.]},
  issn = {0003-1305},
  doi = {10.2307/2684178},
  urldate = {2023-03-06}
}

@article{banaszek2013focus,
  title = {Focus on Quantum Tomography},
  author = {Banaszek, K. and Cramer, M. and Gross, D.},
  year = 2013,
  month = dec,
  journal = {New Journal of Physics},
  volume = {15},
  number = {12},
  pages = {125020},
  publisher = {IOP Publishing},
  issn = {1367-2630},
  doi = {10.1088/1367-2630/15/12/125020},
  urldate = {2023-01-17},
  langid = {english}
}

@article{boixo2018characterizing,
  title = {Characterizing {{Quantum Supremacy}} in {{Near-Term Devices}}},
  author = {Boixo, Sergio and Isakov, Sergei V. and Smelyanskiy, Vadim N. and Babbush, Ryan and Ding, Nan and Jiang, Zhang and Bremner, Michael J. and Martinis, John M. and Neven, Hartmut},
  year = 2018,
  month = jun,
  journal = {Nature Physics},
  volume = {14},
  number = {6},
  eprint = {1608.00263},
  primaryclass = {quant-ph},
  pages = {595--600},
  issn = {1745-2473, 1745-2481},
  doi = {10.1038/s41567-018-0124-x},
  urldate = {2022-12-12},
  archiveprefix = {arXiv}
}

@article{brandao2016local,
  title = {Local {{Random Quantum Circuits}} Are {{Approximate Polynomial-Designs}}},
  author = {Brand{\~a}o, Fernando G. S. L. and Harrow, Aram W. and Horodecki, Micha{\l}},
  year = 2016,
  month = sep,
  journal = {Communications in Mathematical Physics},
  volume = {346},
  number = {2},
  pages = {397--434},
  issn = {0010-3616, 1432-0916},
  doi = {10.1007/s00220-016-2706-8},
  urldate = {2026-01-22},
  langid = {english}
}

@article{bremner2016averagecase,
  title = {Average-{{Case Complexity Versus Approximate Simulation}} of {{Commuting Quantum Computations}}},
  author = {Bremner, Michael J. and Montanaro, Ashley and Shepherd, Dan J.},
  year = 2016,
  month = aug,
  journal = {Physical Review Letters},
  volume = {117},
  number = {8},
  pages = {080501},
  issn = {0031-9007, 1079-7114},
  doi = {10.1103/PhysRevLett.117.080501},
  urldate = {2026-01-22},
  copyright = {http://link.aps.org/licenses/aps-default-license},
  langid = {english}
}

@article{cerezo2021cost,
  title = {Cost Function Dependent Barren Plateaus in Shallow Parametrized Quantum Circuits},
  author = {Cerezo, M. and Sone, Akira and Volkoff, Tyler and Cincio, Lukasz and Coles, Patrick J.},
  year = 2021,
  month = mar,
  journal = {Nature Communications},
  volume = {12},
  number = {1},
  pages = {1791},
  publisher = {Nature Publishing Group},
  issn = {2041-1723},
  doi = {10.1038/s41467-021-21728-w},
  urldate = {2022-12-15},
  copyright = {2021 The Author(s)},
  langid = {english}
}

@misc{chen2024power,
  title = {The Power of a Single {{Haar}} Random State: Constructing and Separating Quantum Pseudorandomness},
  shorttitle = {The Power of a Single {{Haar}} Random State},
  author = {Chen, Boyang and Coladangelo, Andrea and Sattath, Or},
  year = 2024,
  month = oct,
  number = {arXiv:2404.03295},
  eprint = {2404.03295},
  primaryclass = {quant-ph},
  publisher = {arXiv},
  doi = {10.48550/arXiv.2404.03295},
  urldate = {2025-02-26},
  archiveprefix = {arXiv}
}

@article{eisert2020quantum,
  title = {Quantum Certification and Benchmarking},
  author = {Eisert, Jens and Hangleiter, Dominik and Walk, Nathan and Roth, Ingo and Markham, Damian and Parekh, Rhea and Chabaud, Ulysse and Kashefi, Elham},
  year = 2020,
  month = jul,
  journal = {Nature Reviews Physics},
  volume = {2},
  number = {7},
  pages = {382--390},
  publisher = {Nature Publishing Group},
  issn = {2522-5820},
  doi = {10.1038/s42254-020-0186-4},
  urldate = {2025-05-21},
  copyright = {2020 Springer Nature Limited},
  langid = {english}
}

@article{elben2023randomized,
  title = {The Randomized Measurement Toolbox},
  author = {Elben, Andreas and Flammia, Steven T. and Huang, Hsin-Yuan and Kueng, Richard and Preskill, John and Vermersch, Beno{\^i}t and Zoller, Peter},
  year = 2023,
  month = jan,
  journal = {Nature Reviews Physics},
  volume = {5},
  number = {1},
  pages = {9--24},
  publisher = {Nature Publishing Group},
  issn = {2522-5820},
  doi = {10.1038/s42254-022-00535-2},
  urldate = {2023-01-17},
  copyright = {2022 Springer Nature Limited},
  langid = {english}
}

@misc{grassl2009sicpovms,
  title = {On {{SIC-POVMs}} and {{MUBs}} in {{Dimension}} 6},
  author = {Grassl, Markus},
  year = 2009,
  month = may,
  number = {arXiv:quant-ph/0406175},
  eprint = {quant-ph/0406175},
  publisher = {arXiv},
  doi = {10.48550/arXiv.quant-ph/0406175},
  urldate = {2024-04-08},
  archiveprefix = {arXiv}
}

@article{gross2007evenly,
  title = {Evenly Distributed Unitaries: On the Structure of Unitary Designs},
  shorttitle = {Evenly Distributed Unitaries},
  author = {Gross, D. and Audenaert, K. and Eisert, J.},
  year = 2007,
  month = may,
  journal = {Journal of Mathematical Physics},
  volume = {48},
  number = {5},
  eprint = {quant-ph/0611002},
  pages = {052104},
  issn = {0022-2488, 1089-7658},
  doi = {10.1063/1.2716992},
  urldate = {2023-04-21},
  archiveprefix = {arXiv}
}

@article{harrow2023approximate,
  title = {Approximate {{Unitary}} T-{{Designs}} by {{Short Random Quantum Circuits Using Nearest-Neighbor}} and {{Long-Range Gates}}},
  author = {Harrow, Aram W. and Mehraban, Saeed},
  year = 2023,
  month = jul,
  journal = {Communications in Mathematical Physics},
  volume = {401},
  number = {2},
  pages = {1531--1626},
  issn = {1432-0916},
  doi = {10.1007/s00220-023-04675-z},
  urldate = {2025-02-26},
  langid = {english}
}

@article{holmes2022connecting,
  title = {Connecting {{Ansatz Expressibility}} to {{Gradient Magnitudes}} and {{Barren Plateaus}}},
  author = {Holmes, Zo{\"e} and Sharma, Kunal and Cerezo, M. and Coles, Patrick J.},
  year = 2022,
  month = jan,
  journal = {PRX Quantum},
  volume = {3},
  number = {1},
  pages = {010313},
  publisher = {American Physical Society},
  doi = {10.1103/PRXQuantum.3.010313},
  urldate = {2022-12-15}
}

@article{hughes2021spherical,
  title = {Spherical (t,t)-Designs with a Small Number of Vectors},
  author = {Hughes, Daniel and Waldron, Shayne},
  year = 2021,
  month = jan,
  journal = {Linear Algebra and its Applications},
  volume = {608},
  pages = {84--106},
  issn = {0024-3795},
  doi = {10.1016/j.laa.2020.08.010},
  urldate = {2023-06-01},
  langid = {english}
}

@misc{hunter-jones2019unitary,
  title = {Unitary Designs from Statistical Mechanics in Random Quantum Circuits},
  author = {{Hunter-Jones}, Nicholas},
  year = 2019,
  month = may,
  number = {arXiv:1905.12053},
  eprint = {1905.12053},
  primaryclass = {cond-mat, physics:hep-th, physics:quant-ph},
  publisher = {arXiv},
  doi = {10.48550/arXiv.1905.12053},
  urldate = {2024-08-26},
  archiveprefix = {arXiv}
}

@misc{ji2018pseudorandom,
  title = {Pseudorandom {{Quantum States}}},
  author = {Ji, Zhengfeng and Liu, Yi-Kai and Song, Fang},
  year = 2018,
  number = {2018/544},
  url = {https://eprint.iacr.org/2018/544},
  urldate = {2024-05-14}
}

@book{johnson1994continuous,
  title = {Continuous Univariate Distributions. 1},
  author = {Johnson, Norman L. and Kotz, Samuel and Balakrishnan, Narayanaswamy},
  year = 1994,
  edition = {2. ed., 3. [print.] - 1994},
  publisher = {Wiley},
  address = {New York},
  isbn = {978-0-471-58495-7},
  langid = {english}
}

@article{kern2010complete,
  title = {Complete Sets of Cyclic Mutually Unbiased Bases in Even Prime Power Dimensions},
  author = {Kern, Oliver and Ranade, Kedar S. and Seyfarth, Ulrich},
  year = 2010,
  month = jul,
  journal = {Journal of Physics A: Mathematical and Theoretical},
  volume = {43},
  number = {27},
  eprint = {0912.4661},
  primaryclass = {quant-ph},
  pages = {275305},
  issn = {1751-8113, 1751-8121},
  doi = {10.1088/1751-8113/43/27/275305},
  urldate = {2023-02-20},
  archiveprefix = {arXiv}
}

@misc{kueng2015qubit,
  title = {Qubit Stabilizer States Are Complex Projective 3-Designs},
  author = {Kueng, Richard and Gross, David},
  year = 2015,
  month = oct,
  number = {arXiv:1510.02767},
  eprint = {1510.02767},
  primaryclass = {quant-ph},
  publisher = {arXiv},
  doi = {10.48550/arXiv.1510.02767},
  urldate = {2023-10-04},
  archiveprefix = {arXiv}
}

@article{mcclean2018barren,
  title = {Barren Plateaus in Quantum Neural Network Training Landscapes},
  author = {McClean, Jarrod R. and Boixo, Sergio and Smelyanskiy, Vadim N. and Babbush, Ryan and Neven, Hartmut},
  year = 2018,
  month = nov,
  journal = {Nature Communications},
  volume = {9},
  number = {1},
  pages = {4812},
  publisher = {Nature Publishing Group},
  issn = {2041-1723},
  doi = {10.1038/s41467-018-07090-4},
  urldate = {2022-12-08},
  copyright = {2018 The Author(s)},
  langid = {english}
}

@book{meckes2019random,
  title = {The {{Random Matrix Theory}} of the {{Classical Compact Groups}}},
  author = {Meckes, Elizabeth S.},
  year = 2019,
  series = {Cambridge {{Tracts}} in {{Mathematics}}},
  publisher = {Cambridge University Press},
  address = {Cambridge},
  doi = {10.1017/9781108303453},
  urldate = {2024-08-07},
  isbn = {978-1-108-41952-9}
}

@article{perez-salinas2024analyzing,
  title = {Analyzing Variational Quantum Landscapes with Information Content},
  author = {{P{\'e}rez-Salinas}, Adri{\'a}n and Wang, Hao and {Bonet-Monroig}, Xavier},
  year = 2024,
  month = feb,
  journal = {npj Quantum Information},
  volume = {10},
  number = {1},
  eprint = {2303.16893},
  primaryclass = {quant-ph},
  pages = {27},
  issn = {2056-6387},
  doi = {10.1038/s41534-024-00819-8},
  urldate = {2024-08-06},
  archiveprefix = {arXiv}
}

@article{porter1956fluctuations,
  title = {Fluctuations of {{Nuclear Reaction Widths}}},
  author = {Porter, C. E. and Thomas, R. G.},
  year = 1956,
  month = oct,
  journal = {Physical Review},
  volume = {104},
  number = {2},
  pages = {483--491},
  publisher = {American Physical Society},
  doi = {10.1103/PhysRev.104.483},
  urldate = {2024-03-29}
}

@misc{ragone2023unified,
  title = {A {{Unified Theory}} of {{Barren Plateaus}} for {{Deep Parametrized Quantum Circuits}}},
  author = {Ragone, Michael and Bakalov, Bojko N. and Sauvage, Fr{\'e}d{\'e}ric and Kemper, Alexander F. and Marrero, Carlos Ortiz and Larocca, Martin and Cerezo, M.},
  year = 2023,
  month = sep,
  number = {arXiv:2309.09342},
  eprint = {2309.09342},
  primaryclass = {quant-ph},
  publisher = {arXiv},
  doi = {10.48550/arXiv.2309.09342},
  urldate = {2023-11-22},
  archiveprefix = {arXiv}
}

@article{scott2006tight,
  title = {Tight Informationally Complete Quantum Measurements},
  author = {Scott, A. J.},
  year = 2006,
  month = oct,
  journal = {Journal of Physics A: Mathematical and General},
  volume = {39},
  number = {43},
  eprint = {quant-ph/0604049},
  pages = {13507--13530},
  issn = {0305-4470, 1361-6447},
  doi = {10.1088/0305-4470/39/43/009},
  urldate = {2023-06-05},
  archiveprefix = {arXiv}
}

@article{sole1991covering,
  title = {The {{Covering Radius}} of {{Spherical Designs}}},
  author = {Sol{\'e}, Patrick},
  year = 1991,
  month = sep,
  journal = {European Journal of Combinatorics},
  volume = {12},
  number = {5},
  pages = {423--431},
  issn = {0195-6698},
  doi = {10.1016/S0195-6698(13)80018-2},
  urldate = {2023-04-17},
  langid = {english}
}

@article{tavakoli2021mutually,
  title = {Mutually Unbiased Bases and Symmetric Informationally Complete Measurements in {{Bell}} Experiments},
  author = {Tavakoli, Armin and Farkas, M{\'a}t{\'e} and Rosset, Denis and Bancal, Jean-Daniel and Kaniewski, J{\k e}drzej},
  year = 2021,
  month = feb,
  journal = {Science Advances},
  volume = {7},
  number = {7},
  eprint = {1912.03225},
  primaryclass = {quant-ph},
  pages = {eabc3847},
  issn = {2375-2548},
  doi = {10.1126/sciadv.abc3847},
  urldate = {2023-07-12},
  archiveprefix = {arXiv}
}

@article{waldron2003generalized,
  title = {Generalized {{Welch}} Bound Equality Sequences Are Tight Frames},
  author = {Waldron, S.},
  year = 2003,
  month = sep,
  journal = {IEEE Transactions on Information Theory},
  volume = {49},
  number = {9},
  pages = {2307--2309},
  issn = {1557-9654},
  doi = {10.1109/TIT.2003.815788}
}

@article{welch1974lower,
  title = {Lower Bounds on the Maximum Cross Correlation of Signals ({{Corresp}}.)},
  author = {Welch, L.},
  year = 1974,
  month = may,
  journal = {IEEE Transactions on Information Theory},
  volume = {20},
  number = {3},
  pages = {397--399},
  issn = {1557-9654},
  doi = {10.1109/TIT.1974.1055219}
}

@article{zhong2020quantum,
  title = {Quantum Computational Advantage Using Photons},
  author = {Zhong, Han-Sen and Wang, Hui and Deng, Yu-Hao and Chen, Ming-Cheng and Peng, Li-Chao and Luo, Yi-Han and Qin, Jian and Wu, Dian and Ding, Xing and Hu, Yi and Hu, Peng and Yang, Xiao-Yan and Zhang, Wei-Jun and Li, Hao and Li, Yuxuan and Jiang, Xiao and Gan, Lin and Yang, Guangwen and You, Lixing and Wang, Zhen and Li, Li and Liu, Nai-Le and Lu, Chao-Yang and Pan, Jian-Wei},
  year = 2020,
  month = dec,
  journal = {Science},
  volume = {370},
  number = {6523},
  pages = {1460--1463},
  issn = {0036-8075, 1095-9203},
  doi = {10.1126/science.abe8770},
  urldate = {2023-04-03},
  langid = {english}
}

@article{zhu2017multiqubit,
  title = {Multiqubit {{Clifford}} Groups Are Unitary 3-Designs},
  author = {Zhu, Huangjun},
  year = 2017,
  month = dec,
  journal = {Physical Review A},
  volume = {96},
  number = {6},
  eprint = {1510.02619},
  primaryclass = {quant-ph},
  pages = {062336},
  issn = {2469-9926, 2469-9934},
  doi = {10.1103/PhysRevA.96.062336},
  urldate = {2024-12-20},
  archiveprefix = {arXiv}
}

@article{zyczkowski1994random,
  title = {Random Unitary Matrices},
  author = {Zyczkowski, K. and Kus, M.},
  year = 1994,
  month = jun,
  journal = {Journal of Physics A: Mathematical and General},
  volume = {27},
  number = {12},
  pages = {4235},
  issn = {0305-4470},
  doi = {10.1088/0305-4470/27/12/028},
  urldate = {2024-08-07},
  langid = {english}
}
\clearpage

\appendix

\section{Haar-random Discriminator algorithm and subroutines}\label{app.algos}
We summarize our Haar-random discriminator method as pseudo-code algorithms.
The first step is to measure the expectation values of an observable with known spectrum w.r.t. Haar-randomly distributed quantum states.
In parallel, one computes classically the exact moments of the \(t\)-design with \(\Obs\) following~\Cref{eq.exact_moments,le.moments}.
The underlying assumption is then that the random states in the set \(S\) live in a Hilbert space of dimension $N$.
The algorithm described here works for any $t$ without extra measurements.

\begin{algorithm2e}
\caption{Haar-random Discriminator.}\label{alg:discriminator}
\textbf{Procedure:} HaarRandDiscriminator(\(\Obs\), \(S\), \(t_{\rm max}\), \(\mathcal{P}\), \(\mathcal U\))\;
    Set \(\mut(\Obs)\) from~\Cref{eq.exact_moments}\;
    Set \(M\) from~\Cref{eq.montecarlosamples}\;
    \(\mutbar(\Obs, S) \leftarrow\) tDesignCheck(\(\Obs\), \(S\), \(t_{\rm max}\))\;
    Compute \(\avgrand\) from~\Cref{eq:Discriminator}\;
    \If{\(\vert\avgrand\vert > \delta(x)\)}{
        \textbf{Output:} \(S\) is NOT COMPATIBLE with \(\Obs\)-shadowed \(t\)-design\;
        \textbf{STOP}\;
    }
    \Else{
        \(\mutbar(\Obs_{\Pi}, S) \leftarrow\)  PermutationTest(\(\Obs\), \(S\), \(t_{\rm max}\), \(\mathcal{P}\))\;
        Compute \(\avgrandperm\) from~\Cref{eq:Discriminator}\;
        \If{\(\vert\avgrandperm\vert > \delta(x)\)}{
            \textbf{Output:} \(S\) is NOT COMPATIBLE with \(\ObsPi\)-shadowed \(t\)-design\;
            \textbf{STOP}\;
            }
            \Else{
                 \(\mutbar(\Obs_{\Pi,U}, S) \leftarrow\) MUBTest(\(\Obs\), \(S\), \(t_{\rm max}\), \(\mathcal{P}\), \(\mathcal U\))\;
                  Compute \(\avgrandpermU\) from~\Cref{eq:Discriminator}\;
                \If{\(\vert\avgrandpermU\vert > \delta(x)\)}{
                \textbf{Output:} \(S\) is NOT COMPATIBLE with \(\ObsPiU\)-shadowed \(t\)-design\;
                \textbf{STOP}\;}
                \Else{
                \textbf{Output:} \(S\) is COMPATIBLE with a \(t\)-design\;
                }
        }
    }   
\end{algorithm2e}

\begin{algorithm2e}
\caption{\(\Obs\)-shadow \(t\)-design}\label{alg:tdesigntest}
\textbf{Procedure:} tDesignCheck(\(\Obs\), \(S\), \(t_{\rm max}\))\;
\textbf{Input:} a quantum observable \(\Obs\), a black-box quantum device that outputs quantum states \(\ket{\Psi_i} \in S\), maximum \(t\)-moment to estimate \(t_{\rm max}\)\;
    Sample \(M\) quantum states from \(S\)\;
    \For{\(i \in [1...M]\)}{
    Measure \(\ExpObs[\Obs]_i = \bra{\Psi_i} \Obs \ket{\Psi_i}\)\;
    }
    \For{\(t \in [1...t_{\rm max}]\)}{
    Estimate \(\mutbar(\Obs, S)\) from~\Cref{eq:mutbar}\;
    }
\textbf{Output:} \(\mutbar(\Obs, S)\)
\end{algorithm2e}

\begin{algorithm2e}
\caption{Add Permutations}\label{alg:permtest}
\textbf{Procedure:} PermutationCheck(\(\Obs\), \(S\), \(t_{\rm max}\), \(\mathcal{P}\))\;
\textbf{Input:} a quantum observable \(\Obs\), a black-box quantum device that outputs quantum states \(\ket{\Psi_i} \in S\), maximum \(t\)-moment to estimate \(t_{\rm max} > 1\), and a set of permutations \(\mathcal{P}\)\;
    \For{\(\Pi \in \mathcal{P}\)}{
    Set \(\Obs_{\Pi} = \Pi \Obs \Pi\)\;
    Call tDesignTest(\(\Obs_{\Pi}\), \(S\), \(t_{\rm max}\))\;
    }
\textbf{Output:}  \(\mutbar(\Obs_{\Pi}, S)\)
\end{algorithm2e}

\begin{algorithm2e}
\caption{Add Mutually Unbiased Bases}\label{alg:mubtest}
\textbf{Procedure:} MUBCheck(\(\Obs\), \(S\), \(t_{\rm max}\), \(\mathcal{P}\), \(\mathcal U\))\;
\textbf{Input:} a quantum observable \(\Obs\), a black-box quantum device that outputs quantum states \(\ket{\Psi_i} \in S\), maximum \(t\)-moment to estimate \(t_{\rm max} > 1\), a set of permutations \(\mathcal{P}\), and a complete set of Mutually Unbiased Bases \(\mathcal U\)\;
    \For{\(U \in \mathcal{U}\)}{
    Set \(\Obs_{U} = U^{\dagger} \Obs U\)\;
    Call PermutationTest(\(\Obs_{U}\), \(S\), \(t_{\rm max}\), \(\mathcal{P}\))\;
    }
\textbf{Output:}  \(\mutbar(\Obs_{\Pi, U}, S)\)

\end{algorithm2e}

\newpage

\section{Proof of~\Cref{le.dirichlet} and~\Cref{th.dirichlet}}\label{app.dirichlet}

Consider a set of states $\{\ket\psi\} \sim \H$, where \H~is the Haar-random distribution over the complex projective space $\mathbb C P^{N - 1}$, defined as
\begin{equation}
    S^N = \{\ket\psi \colon \ket\psi \in \mathbb C^N, \Vert \ket\psi \Vert = 1 \}.
\end{equation}

We begin by defining the Dirichlet distribution.
\begin{definition}[Dirichlet distribution~\cite{johnson1994continuous}]\label{def.dirichlet}
The Dirichlet distribution $\bm x \sim \Dirichlet$ parameterized by $\alpha\in\mathbb{R}_{>0}^N$ is supported on the $(N-1)$-standard simplex, i.e., $\bm x=(x_1, x_2, \ldots, x_N), \Vert \bm x \Vert_1 = 1$. It has the following probability density function with respect to the Lebesgue measure on $\mathbb{R}^{N-1}$: 
\begin{equation}
    f_{\rm Dir}(\bm x, \bm \alpha) = \frac{\Gamma(\alpha_0)}{\prod_{i = 1}^N\Gamma(\alpha_i)}\prod_{i = 1}^N x_i^{\alpha_i - 1},
\end{equation}
where $\alpha_0 = \Vert \bm\alpha\Vert_1$.
It is possible to exactly compute the statistical moments of arbitrary order $\bm k = (k_1, k_2, \ldots, k_N)$,
\begin{equation}\label{eq.dir_moments}
    \E[\bm x \sim \Dirichlet]{\prod_{i = 1}^N x_i^{k_i}} = \frac{\Gamma(\alpha_0)}{\Gamma(\alpha_0 + k_0)} \prod_{i = 1}^N \frac{\Gamma(\alpha_i + k_i)}{\Gamma(\alpha_i)}.
\end{equation}
\end{definition}

We now show the relationship between Haar-random states and the Dirichlet distribution.
By assumption, a Haar-random ensemble $\{\ket\psi \}$ and $\{ U \ket\psi\}$ are statistically equivalent, for any unitary matrix $U \in \mathcal{SU}(N)$. The projections $\vert \bra\psi U \ket i\vert^2$ are now a set of random variables, subject to
\begin{equation}
    \sum_{i = 0}^{N - 1} \vert \bra\psi U \ket i\vert^2 = 1.
\end{equation}
Therefore, the random variables $x_i = \vert \bra\psi U \ket i\vert^2$ must follow a Dirichlet distribution. This must hold for any $U$. In particular, it must hold for any permutation between elements in the computational basis, and therefore the random variables $x_i$ and $x_j$ must be indistinguishable for any pair $(i, j)$. This yields 
\begin{equation}
    \bm x \sim \Dirichlet[\alpha \bm 1],
\end{equation}
where $\bm 1$ is a vector with all entries equal to 1, and $\alpha$ is a normalization constant, and it is the only missing part. For the normalization, we can recall the known result that coordinates (squared) in a multidimensional sphere can be expressed as Dirichlet distributions with parameters $1/2$~\cite{bailey1992distributional}. Notice that our quantum states have $N$ independent complex values, thus each coordinate has the double contribution. The additivity of Dirichlet distribution over its coordinates suffices to state $\alpha = 1/2 + 1/2 = 1$. This completes the proof. \qed

We proceed now to proof~\Cref{th.dirichlet}.
The sets of quantum states are Haar-random. We choose $U$ as the matrix that diagonalizes the observable $\Obs$. This allows us to express the quantity $\bra\psi \Obs \ket\psi$ in the basis of $\Obs$ as
\begin{equation}\label{eq.sum_dir}
    \bra\psi \Obs \ket\psi = \sum_{i = 1}^N \lambda_i \vert \braket{\lambda_i}{\psi}\vert^2,
\end{equation}
where $\lambda_i$ are the eigenvalues of $H$. Trivially, we obtain that 
\begin{equation}
    \bra\psi \Obs \ket\psi = \sum_{i = 1}^{2^n} \lambda_i x_i, 
\end{equation}
with $\bm x \sim \Dirichlet[\bm 1 / 2]$.
 
Notice that the symmetric Dirichlet distribution implies that $\braketObs$ is insensitive to any permutation of coordinates $(i, j)$ if $\lambda_i = \lambda_j$. This can be related to the aggregation property of the Dirichlet distribution, namely
\begin{equation}\label{eq.aggregation}
    (x_1, x_2, \ldots, x_N) \sim \Dirichlet[\alpha_1, \alpha_2, \ldots, \alpha_N] \rightarrow
    (x_1, x_2, \ldots, x_i + x_j, \ldots) \sim \Dirichlet[\alpha_1, \alpha_2, \ldots, \alpha_i + \alpha_j, \ldots].
\end{equation}
Therefore, we can group all coordinates with the same eigenvalue and aggregate them under a single random variable, with the corresponding parameter proportional to the multiplicity of the eigenvalue. This yields the desired result.
\qed

\section{Proof of~\Cref{le.moments}}\label{app.moments}

The proof is as follows. In order to compute $\mut(\Obs)$ we recall the multinomial theorem as 
\begin{equation}\label{eq.multinomial}
    \E[\Dirichlet]{\left(\sum_{i = G}\lambda_i x_i\right)^t} = \sum_{\substack{\bm k \in \mathbb N^G \\ \Vert \bm k \Vert_1 = t}} \binom{t}{\bm k} \left(\prod_{i=1}^G \lambda_i^{k_i}\right) \E[\Dirichlet]{\prod_{i=1}^G x_i^{k_i}}, 
\end{equation}
where the multinomial coefficient is defined as
\begin{equation}
    \binom{t}{\bm k} = \frac{t!}{\prod_{i = 1}^G k_i!}.
\end{equation}

We recall now~\Cref{eq.dir_moments} for computing the expectation values over products of $x_i$. This yields the exact formula
\begin{equation}\label{eq:th_mut}
    \E[\Dirichlet]{\left(\sum_{i = G}\lambda_i x_i\right)^t} = \sum_{\substack{\bm k \in \mathbb N^G \\ \Vert \bm k \Vert_1 = t}} \binom{t}{\bm k} \left(\prod_{i=1}^G \lambda_i^{k_i}\right) \frac{\Gamma(\alpha_0)}{\Gamma(\alpha_0 + t)} \prod_{i = 1}^N \frac{\Gamma(\alpha_i + k_i)}{\Gamma(\alpha_i)}, 
\end{equation}
with $\alpha_0 = \Vert \bm \alpha \Vert_1$.
\qed

\section{Proof of~\Cref{cor.efficient_moments}}\label{app.efficient_moments}
For the proof, we resume the proof of~\Cref{le.moments}. As a first approximation, we recall 
\begin{equation}
    \Gamma(x + a) \approx x^a \Gamma(x).
\end{equation}
This simple approximation allows us to write
\begin{equation}
    \E[\Dirichlet]{\left(\sum_{i = G}\lambda_i x_i\right)^t} \approx \frac{1}{\alpha_0^t}\sum_{\substack{\bm k \in \mathbb N^G \\ \Vert \bm k \Vert_1 = t}} \binom{t}{\bm k} \prod_{i = 1}^G (\lambda_i \alpha_i)^{k_i}. 
\end{equation}
By definition, the terms in the sum match the multinomial theorem for $\alpha_i = m_i$. This yields the easy approximation
\begin{equation}
     \E[\Dirichlet]{\left(\sum_{i = G}\lambda_i x_i\right)^t} \approx \left( \frac{\Tr(\Obs)}{N}\right)^t.
\end{equation}

This first approximation implies that $\mut(\Obs) \approx \mu_1(\Obs)^t$, and it is not enough to carry out faithful comparisons. To this end, we simplify the statistical moments by applying Stirling's approximation with more terms.
\begin{equation}
    \log \Gamma(x) = x \log x - x + \mathcal O(\log(x)), 
\end{equation}
thus
\begin{equation}
    \log\left(\frac{\Gamma(\alpha + t)}{\Gamma(\alpha)}\right) = (\alpha + t)\log\left(\alpha + t\right) - \alpha \log(\alpha) - t. 
\end{equation}

Then 
\begin{align}
    \log\left( \E[{\Dirichlet}]{\prod_{i=1}^G x_i^{k_i}}\right) = & \log\left( \frac{\Gamma(\alpha_0)}{\Gamma(\alpha_0 + t)}\right) + \sum_{i=1}^G \log\left(\frac{\Gamma(\alpha_i + k_i)}{\alpha_i} \right) = \\ 
    & \alpha_0 \log{\alpha_0} + t - \left(\alpha_0 + t \right) \log\left(\alpha_0 + t\right)
    + \mathcal O \log(\alpha + t)\nonumber \\ 
    & \sum_{i = 1}^G - \alpha_i \log{\alpha_i} - k_i + \left(\alpha_i + k_i \right) \log\left(\alpha_i + k_i\right)= \\
     \alpha_0 \log{\alpha_0} + t - \left(\alpha_0 + t \right) \log\left(\alpha_0 + t\right) &- \sum_{i = 1}^G \alpha_i \log{\alpha_i} - \left(\alpha_i + k_i \right) \log\left(\alpha_i + k_i\right) + \mathcal O \log(\alpha + t)
\end{align}

Therefore, neglecting the errors, 
\begin{equation}\label{eq.after_stirling}
    \E[{\Dirichlet}]{\prod_{i=1}^G x_i^{k_i}} = \frac{\alpha_0^{\alpha_0}}{(\alpha_0 + t)^{\alpha_0 + t}} \prod_{i = 1}^G \frac{(\alpha_i + k_i)^{\alpha_i + k_i}}{\alpha_i^{\alpha_i}}
\end{equation}

We now pull the bounds
\begin{align}
    \alpha_i^{k_i} \exp\left(k_i\right) \leq & \frac{(\alpha_i + k_i)^{\alpha_i + k_i}}{\alpha_i^{\alpha_i}} \leq \alpha_i^{k_i} \exp\left(k_i\left(1+\frac{k_i}{2\alpha_i}\right)\right) \\
    \alpha_0^{-t} \exp\left(-t\left(1 + \frac{t}{2\alpha_0}\right)\right) \leq & \frac{\alpha_0^{\alpha_0}}{(\alpha_0 + t)^{\alpha_0 + t}} \leq \alpha_0^{-t} \exp(-t)
\end{align}
We now plug these expansions into~\Cref{eq.after_stirling}, again dropping the error, to obtain 
\begin{equation}
    \exp\left(-\frac{t^2}{2\alpha}\right) \leq \frac{\E[{\Dirichlet}]{\prod_{i=1}^G x_i^{k_i}}}{\alpha_0^{-t}\prod_{i = 1}^G \alpha_i^{k_i}} \leq
    \exp\left(\sum_{i = 1}^G \frac{k^2_i}{2\alpha_i}\right)
\end{equation}
Now, we recall \Cref{eq:th_mut} to rewrite it as
\begin{equation}
    \E[\Dirichlet]{\left(\sum_{i = G}\lambda_i x_i\right)^t} = \sum_{\substack{\bm k \in \mathbb N^G \\ \Vert \bm k \Vert_1 = t}} \binom{t}{\bm k} \left(\prod_{i=1}^G \lambda_i^{k_i}\right) \alpha_0^{-t} \prod_{i = 1}^N \alpha_i^{k_i} f(\bm k_i, t),  
\end{equation}
with 
\begin{equation}
   \exp\left(-\frac{-t^2}{2N} \right)\leq f(\bm k_i, t) \leq \exp\left(\sum_{i = 1}^G\frac{k_i^2}{2 \alpha_i}\right)
\end{equation}
These bounds give a multiplicative factor difference with respect to the averages, which injected into the calculations for $\mut(\Obs)$ yields
\begin{align}
    \mut(\Obs) & \leq \left(\frac{\Tr \Obs}{N}\right)^t \exp\left(\frac{t^2}{2 \bar\alpha}\right) \\
    \mut(\Obs) & \geq \left(\frac{\Tr \Obs}{N}\right)^t\exp\left(-\frac{t^2}{N}\right),
\end{align} 
with $\frac{1}{\bar\alpha} = \sum_{i = 1}^G \alpha^{-1}$.

\qed

\section{Proof of~\Cref{th.Haar_check}}\label{app.Haar_check}
We start by considering the ensemble of states $S = \{\ket\phi\}$, with size $M$.
By Monte Carlo method we obtain the estimate $\mutbar(\Obs, S)$ with error 
\begin{equation}
    \delta(\bar\mu_t(\Obs, S)) = \frac{\hat\sigma_t(\Obs, S)}{\sqrt M}.
\end{equation}
We compare now the estimate $\bar\mu_t(\Obs, S)$ to the Haar-random theoretical value $\mu(\Obs)$ available in~\Cref{le.moments}, or~\Cref{cor.efficient_moments}. This quantity is designed as
\begin{equation}
    \avgrand = \left\vert \bar\mu_t(\Obs, S) - \mu_t(\Obs)\right\vert. 
\end{equation}
Assuming that $\Delta_t$ is small, we can argue that $S$ forms a $t$-design w.r.t. the observable $\Obs$. Due to the Monte Carlo error, this claim is only true with a certain confidence, which depends on the variance $\hat\sigma_t(\Obs, S)$, to be computed numerically. In the affirmative case of $t$-designs, we can estimate the variance easily as
\begin{equation}
    \sigma^2_t(\Obs) = \mu_t(\Obs^2) - \mu_t(\Obs)^2 = \mu_{2t}(\Obs) - \mu_t(\Obs)^2. 
\end{equation}
This variance can be exactly computed from~\Cref{le.moments}. We can easily upper-bound it through~\Cref{cor.efficient_moments} and obtain
\begin{equation}
    \sigma^2_t(\Obs) \leq \left( \frac{\Tr(\Obs)}{N}\right)^{2t} \exp\left( \frac{2t^2}{\bar m}\right), 
\end{equation}
which yields the condition
\begin{equation}
    \delta^2\left(\avgrand \right)\leq \left( \frac{\Tr(\Obs)}{N}\right)^{2t} \frac{\exp\left( \frac{2t^2}{\bar m}\right)}{M}, 
\end{equation}
Therefore, we can only ensure $\epsilon$ accuracy in detecting $t$-designs if the number of samples scales as
\begin{equation}
    M \geq \frac{\exp\left( \frac{t^2}{\bar m}\right)}{\epsilon^2} \left( \frac{\Tr(\Obs)}{N}\right)^{2t}
\end{equation}
\qed

\section{Proof of~\Cref{prop.nonsymmetric}}\label{app.nonsymmetric}

To prove this result, we need to state first some auxiliary elements based on the Dirichlet distribution. 

\begin{lemma}[Almost symmetric Dirichlet distribution]\label{le.almost_symmetric_dirichlet}
    Let $\bm x$ be a Dirichlet-random variable stemming from a Dirichlet distribution with parameters 
    \begin{equation}
        \bm\alpha = \frac{\alpha_0}{N} (\bm 1 + \bm\epsilon), 
    \end{equation}
    with $\sum_{i = 0}^N \epsilon_i = 0$. 
    Then 
    \begin{align}
        \left\vert\E{(\bm\lambda \cdot \bm x)^t} - \E{(\bm\lambda \cdot \bm x_0)^t} \right\vert & \leq \left(\frac{\Tr(\Obs)}{N}\right)^t t \Vert \bm\epsilon \Vert_\infty, \\
        \left\vert\E{(\bm\lambda \cdot \bm x)^t} - \E{(\bm\lambda \cdot \bm x_0)^t} \right\vert & \leq \left(\frac{\Vert \Obs \Vert_\infty}{N}\right)^t t \Vert \bm\epsilon \Vert_1, \\
    \end{align}
    with $\bm x_0$ being the Dirichlet random variable corresponding to $\bm\epsilon = \bm 0$. 
\end{lemma}

To show this, we just need to, again, apply the multinomial theorem to the expectation value, to obtain
\begin{equation}
    \E[\Dirichlet]{\left(\sum_{i = N}\lambda_i x_i\right)^t} = \sum_{\substack{\bm k \in \mathbb N^G \\ \Vert \bm k \Vert_1 = t}} \binom{t}{\bm k} \left(\prod_{i=1}^N \lambda_i^{k_i}\right) \frac{\Gamma(\alpha_0)}{\Gamma(\alpha_0 + t)} \prod_{i = 1}^N \frac{\Gamma(\alpha + \epsilon_i + k_i)}{\Gamma(\alpha + \epsilon_i)}, 
\end{equation}
with $\alpha = \alpha_0 / N$. We focus on the last term and find bounds, that is
$
    \frac{\Gamma(\alpha + \epsilon + k)}{\Gamma(\alpha + \epsilon)}.
$

To bound this element, we can use a Taylor expansion. For small values of $\epsilon$,  
\begin{equation}
    \frac{\Gamma(\alpha (1 + \epsilon) + k)}{\Gamma(\alpha (1 + \epsilon))} = \left( \frac{\Gamma(\alpha + k)}{\Gamma(\alpha)}\right)\left( 1 + \epsilon( \varphi(\alpha + k) - \varphi(\alpha)) + \mathcal{O}(\epsilon^2)\right), 
\end{equation}
where $\varphi(z)$ is the logarithmic derivative of the Gamma function, that is, 
\begin{equation}
    \varphi(x) = \frac{d}{dz} \log(\Gamma(z)) = \frac{\Gamma'(z)}{\Gamma(z)}.
\end{equation}
This function has the property that
\begin{equation}
    \varphi(1 + z) = \varphi(z) + \frac{1}{z}, 
\end{equation}
therefore
\begin{equation}
    \varphi(\alpha + k) - \varphi(\alpha) = \sum_{j = 0}^{k - 1} \frac{1}{\alpha + j} \equiv \frac{C_{k, \alpha}}{\alpha}.
\end{equation}
For $\epsilon \leq 0$, this constant suffices to have an upper-bound, and for $k = 1$, the approximation is an equality. In the case of $\epsilon > 0$, and $k \geq 2$, there exists another constant satisfying a linear upper bound. We plug this result into the analytical calculation.
\begin{multline}
     \E[\Dirichlet]{\left(\sum_{i = 1 }^N \lambda_i x_i\right)^t}  = \\ \sum_{\substack{\bm k \in \mathbb N^N \\ \Vert \bm k \Vert_1 = t}} \binom{t}{\bm k} \left(\prod_{i=1}^N \lambda_i^{k_i}\right) \frac{\Gamma(\alpha_0)}{\Gamma(\alpha_0 + t)} \prod_{i = 1}^N \frac{\Gamma(\alpha + k_i)}{\Gamma(\alpha)}\left(1 + C_{k, \alpha} \epsilon_i + \mathcal O(\epsilon_i^2) \right), 
\end{multline}
and obtain, through Hölder's inequality, 
\begin{equation}
\left\vert\E[\Dirichlet]{(\bm\lambda \cdot \bm x)^t} - \E[\Dirichlet]{(\bm\lambda \cdot \bm x_0)^t} \right\vert \leq \E{(\bm\lambda \cdot \bm x_0)^t} \biggl(C_{t, \alpha}\Vert\bm\epsilon\Vert_\infty + \mathcal O(\Vert \bm\epsilon\Vert^2_\infty)\biggr),
\end{equation}
with $C_{t, \alpha} \in \mathcal O(t)$. Alternatively, 
\begin{equation}
    \left\vert\E[\Dirichlet]{(\bm\lambda \cdot \bm x)^t} - \E[\Dirichlet]{(\bm\lambda \cdot \bm x_0)^t} \right\vert \leq \frac{\Vert \Obs \Vert_\infty^t}{(N\alpha)^t} \frac{\Gamma(\alpha + t)}{\Gamma(\alpha)} C_{t, \alpha} \Vert \bm\epsilon \Vert_1
\end{equation}

\qed

To arrive to \Cref{prop.nonsymmetric}, we can rely on the previous results. We will be interested in computing

\begin{multline}
    \Var\biggl(\E[\Dirichlet]{(\bm\lambda\cdot \bm x)^t} - \E[{\Dirichlet[\alpha \bm 1]}]{(\bm\lambda\cdot \bm x)^t}\biggr) = \\ 
    \Var_{\Pi}\left(\sum_{\substack{\bm k \in \mathbb N^N \\ \Vert \bm k \Vert_1 = t}} \binom{t}{\bm k} \left(\prod_{i=1}^N \lambda_i^{k_i}\right) \frac{\Gamma(\alpha_0)}{\Gamma(\alpha_0 + t)} \prod_{i = 1}^N \frac{\Gamma(\alpha + k_i)}{\Gamma(\alpha)} C_{k, \alpha} \epsilon_{\Pi(i)} + \mathcal O(\epsilon_{\Pi(i)}^2) \right).
\end{multline}

The variable $\epsilon_{\Pi(i)}$ encodes the differences between the different coordinates of $\bm\epsilon$ with respect to the average values. We consider this a random variable which is equal for all $i$, due to symmetry of permutations, and uncorrelated, that is
\begin{equation}
    \Cov_\Pi\left(\epsilon_{\Pi(i), \Pi(j)}\right) = 0, 
\end{equation}
since for every permutation there will be another one with the opposite effect. Hence, 
\begin{multline}
    \Var\biggl(\E[\Dirichlet]{(\bm\lambda\cdot \bm x)^t} - \E[{\Dirichlet[\alpha \bm 1]}]{(\bm\lambda\cdot \bm x)^t}\biggr) = \\ 
    \Var(\epsilon_i) \left(\sum_{\substack{\bm k \in \mathbb N^N \\ \Vert \bm k \Vert_1 = t}} \binom{t}{\bm k}^2 \left(\prod_{i=1}^N \lambda_i^{2k_i}\right) \left(\frac{\Gamma(\alpha_0)}{\Gamma(\alpha_0 + t)}\right)^2 \prod_{i = 1}^N \left(\frac{\Gamma(\alpha + k_i)}{\Gamma(\alpha)}\right)^2 C^2_{k, \alpha} \right). 
\end{multline}
We can provide lower bounds by saying $C_{k, \alpha} \geq 1$ for $k\geq 1$, and $0$ otherwise. For small values of $t$, typical contributions are for $k = 1$, and only $t / N$ of them appear, with constant multiplicity. On the other hand, $\Gamma(\alpha_0 + t) \Gamma(\alpha)^{-1} \approx (\alpha N)^t$, and $\Gamma(\alpha + 1) = \alpha \Gamma(\alpha)$. On the multinomial side, any multinomial factor is smaller than its square. Again making the connection with the multinomial theorem, 
\begin{equation}
    \Var\biggl(\E[\Dirichlet]{(\bm\lambda\cdot \bm x)^t} - \E[{\Dirichlet[\alpha \bm 1]}]{(\bm\lambda\cdot \bm x)^t}\biggr) \in \tilde\Omega\left( \left( \frac{\Tr\Obs^2}{N^2}\right)^t\right)\Var(\epsilon).
\end{equation}
The immediate interpretation is that
\begin{equation}
    \E[i, j]{(\alpha_i - \alpha_j)^2} \in \mathcal O\left( \left(\frac{\Tr(\Obs^2)}{N^2}\right)^{-t} \Var_{\Pi}\left( \mu_t(\ObsPi, S)\right)\right). 
\end{equation}
Notice that $\mut(\Obs)$ scales at first order as $(\Tr \Obs/N)^t$.

To arrive to the final result, we just need to show that the average with respect to permutations is similar to the average of the true symmetric dirichlet distribution. The following corollary supports this claim. 

\begin{corollary}[Average over permutation is close to true Dirichlet]\label{cor.hoeffding_dirichlet}
Let $\bm\epsilon$ and $\bm x$ be defined as in \Cref{le.almost_symmetric_dirichlet}. Let $\Pi \in \mathcal P$ be the set of all possible permutations of the elements of $\bm\epsilon$. Then, 
    \begin{multline}
        \operatorname{Prob}\biggl(\left\vert\E[\Pi \sim \mathcal P]{\E[\Dirichlet]{(\bm\lambda \cdot \bm x)^t}} - \E[\Dirichlet]{(\bm\lambda \cdot \bm x_0)^t}\right\vert \geq \Delta \E[\Dirichlet]{(\bm\lambda \cdot \bm x_0)^t} \biggr) \leq \\ 2 \exp\left( -\frac{2^{2N} \Delta^2 }{C_{t, \alpha}\Vert \bm\epsilon\Vert_\infty^2}\right)
    \end{multline}
\end{corollary}

The proof is an immediate consequence of Hoeffding's inequality and \Cref{le.almost_symmetric_dirichlet}. We put an upper-bound in the value of the differences between the non-symmetric distributions of $\bm x$ that does not depend on the value of $\bm\epsilon$, or the permutation. By applying Hoeffding's inequality, and knowing that the number of possible permutations is $2^N$ we can infer that the average with respect to permutations is almost indistinguishable from the true symmetric Dirichlet distributions, for any expectation values, that is $\bm\epsilon = \bm 0$
.

\qed

To our level of precision, the average over all permutations and the average of true Dirichlet is then indistinguishable, as long as the perturbation is moderate. This yields the final result. 
\qed

\section{Proof of~\Cref{le.ic_povm}}\label{app.ic_povm}

To obtain the full tomographic characterization of a quantum state, one needs to conduct experiments over an informationally complete set of positive-operator valued measurements (IC-POVM). Such sets of operators can perfectly distinguish between two different states, providing a sufficient number of copies. While infinitely many ways exist to construct sets of IC-POVM, some of these sets are larger than others. In particular, there exist recipes to construct tight IC-POVM, i.e., sets of operators with a minimal number of projectors. the number of elements of such tight systems scale as $N^2$. 

An almost tight set of IC-POVM can be obtained by combining a reference basis with a set of mutually unbiased bases (MUB).
Two bases $U, V$ are MUB if
\begin{equation}
    \left\vert \bra i V^\dagger U \ket j \right\vert^2 = \frac{1}{N}, 
\end{equation}
where $N$ is the dimensionality of the system. A complete set of MUB is a set of bases that are pairwise unbiased. The number of MUB in a complete set scales as $(N + 1)$. The existence of a complete set of MUB is an open problem in general~\cite{grassl2009sicpovms}. However, for prime-power dimensions $N = p^n$, it is possible to construct complete sets of MUB~\cite{appleby2014galois, tavakoli2021mutually}.    

Assume now the set of observables $\ket i \bra i$, for any reference basis, are available to measure. Then the set
\begin{equation}
    M_{U, i} = U\ket i \bra i U^\dagger, 
\end{equation}
where $U$ runs over a complete set of MUB is a set of IC-POVMs~\cite{scott2006tight}. 

The only step missing is constructing all measurements corresponding to a single basis through the permutations. 
We consider the observable $\Obs$ in the diagonal basis of the form
\begin{equation}
    \Obs = \operatorname{diag}\left(\underbrace{0, \ldots, 0}_{m_0}, \underbrace{\lambda_1, \ldots, \lambda_1}_{m_1}, \ldots \underbrace{\lambda_G, \ldots, \lambda_G}_{m_G},  \right). 
\end{equation}
We want to construct operators of the form $\Vert \Obs \Vert \left(\ket i \bra i - \ket j \bra j\right)$, with $\Vert \Obs \Vert = \lambda_G$. If $\ket i$ correspond to the $0$-th eigenspace, and $\ket j$ corresponds to the $G$-th eigenspace, then
\begin{equation}
    \Vert \Obs \Vert \left(\ket i \bra i - \ket j \bra j\right) = \Obs - \Pi(i \leftrightarrow j)\ \Obs \ \Pi(i \leftrightarrow j). 
\end{equation}

The prefactor of $\Vert \Obs\Vert$ is independent of the eigenstates $(i, j)$ because the permutations allow the coefficients to be freely moved. In the case $\ket i, \ket j$ belong to eigenspaces in the middle zone of the spectrum, we just need to apply an extra permutation exchanging $\ket i$ with $\ket 0$, and $\ket j$ with $\ket{N - 1}$. 
The new set of operators is given by 
\begin{equation}
    M_{i, j} = \ket i\bra i  - \ket j \bra j. 
\end{equation}
These sets of operations, together with the resolution of the identity 
\begin{equation}
    I = \sum_{i = 0}^{N - 1} \ket i \bra i
\end{equation}
allows us to compute any observable of the form
\begin{equation}
    M_i = \ket i \bra i.
\end{equation}
Together with the sets of MUB~\cite{scott2006tight}, this leads to a set of IC-POVMs. \qed

\end{document}